\documentstyle[12pt]{article}

\newcommand{\newc}{\newcommand}
\newc{\sm}{ the Standard Model }
\newc{\gsim}{\lower.7ex\hbox{$\;\stackrel{\textstyle>}{\sim}\;$} } \newc{\lsim}{\lower.7ex\hbox{$\;\stackrel{\textstyle<}{\sim}\;$}} 
\def\text#1{\hbox{#1}}

\newcommand{\ol}{\overline}

\def\bar#1{{\ol{#1}}}
\def\be{\begin{equation}}

\def\ee{\end{equation}}
\def\bea{\begin{eqnarray}}
\def\eea{\end{eqnarray}}

\begin{document}

\author{Witold Pokorski and Graham G. Ross, \\
Dept. of Physics, \\
Theoretical Physics, \\
University of Oxford, \\
Oxford OX1 3NP.}
\title{Flat directions, String Compactification and 3 Generation Models.}
\date{}
\maketitle

\begin{abstract}
We show how identification of absolutely flat directions allows the
construction of a new class of compactified string theories with reduced
gauge symmetry that may or may not be continuously connected to the original
theory. We use this technique to construct a class of 3 generation models
with just the Standard Model gauge group after compactification. We discuss
the low-energy symmetries necessary for a phenomenologically viable
low-energy model and construct an example in which these symmetries are
identified with string symmetries which remain unbroken down to the
supersymmetry breaking scale. Remarkably the same symmetry responsible for
stabilising the nucleon is also responsible for ensuring one and only one
pair of Higgs doublets is kept light. We show how the string symmetries also
lead to textures in the quark and lepton mass matrices which can explain the
hierarchy of fermion masses and mixing angles.
\end{abstract}

\section{Introduction}

The construction of the effective low-energy theory from the string has
undergone several iterations since the discovery of the heterotic string
opened the way to building realistic 4-D theories with the Standard Model
gauge group. The first attempts compactified the 10-D string theory using
Calabi-Yau manifolds \cite{CY} or orbifolds \cite{orbifolds} and examples of
three generation models were identified \cite{tianyau}. Subsequently it was
realised that one could directly construct 4-D models, many of which did not
have a ten-dimensional interpretation, and so there was no absolute need to
interpret our low-energy world as resulting from a higher dimensional theory.
Recently, in order to reconcile the string prediction for the gauge coupling
unification scale with the ``experimental'' value, it has been
suggested \cite
{witten} that the relevant underlying string theory is actually the strongly
interacting heterotic string, or M-theory \cite{horavawitten}. Many of the
4-D string theory constructions cannot be embedded in M-theory and as a
result the discussion has come full circle and there is renewed interest in
the construction compactified string theories. Although the starting point
is eleven dimensional, much of the 10-D work remains relevant because the
strongly coupled heterotic string theory corresponds, in leading order in
the string coupling, to M-theory compactified on the direct product of a 6-D
Calabi-Yau manifold with the line interval, i.e. $CY_{3}\otimes S_{1}/Z_{2}.$

In order to construct the effective low-energy theory descending from
M-theory which directly realises the prediction of the string for the string
unification scale, $M_{X},$ one must construct the compactified theory on a
specific $CY_{3}\otimes S_{1}/Z_{2}$ manifold such that the gauge group
after compactification should be just $SU(3)\otimes SU(2)\otimes U(1)$,
otherwise the prediction of the string relating gauge couplings will not
refer to just those couplings measured in the laboratory but rather those of
the Grand Unified group. This means there should be {\it no} GUT below the
compactification scale! In this case one may ask why then does the Standard
model have GUT\ properties? This need not be a difficult question to answer
because the string {\it does} have an underlying GUT\ structure. In this
case GUT\ breaking on compactification \ must leave some relic of an
underlying string gauge symmetry in the low energy spectrum. It is the
purpose of this paper to investigate in detail the structure to be expected
in such cases by constructing explicit three generation string models
compactified on Calabi Yau manifolds.

To date, the three generation Calabi-Yau manifolds constructed have all had
a gauge group larger than that of the Standard Model after compactification.
The reason is straightforward. The models are based on the heterotic string
with an underlying $E_{8}\otimes E_{6}$ gauge symmetry. In order to break
this group at the compactification scale it is necessary to implement Wilson
line breaking on a non-simply-connected manifold. The construction of the
three generation case involves the modding out of the simply connected
manifold by a discrete symmetry group and it is this process that introduces
the non-simply- connectedness of the manifold. The associated Wilson lines
form a representation of some or all of the discrete symmetry group.
However, since the discrete groups involved are $Z_{3}$ groups, the Wilson
line breaking is limited and, at best, can only reduce the gauge group to $%
SU(3)^{3} \cite{segreovrut}$. However it is known that these
compactifications, which have a (2,2) world sheet supersymmetry, are related
along flat directions to theories with (2,0) supersymmetry \cite{gsw}. The
latter have a reduced symmetry $E_{8}\otimes SO(10)$ or $E_{8}\otimes SU(5)$
and thus, after Wilson line breaking, may provide a basis for constructing a
three generation model with just the Standard model gauge group. However not
much is known about these (2,0) models for they have proved difficult to
construct. Some examples are known \cite{greene}, built using the direct
product of superconformal theories constrained to have the correct ghost
charge, but none with just three generations. In this paper we will develop
a method which does allow us to construct three generation examples with
(2,0) symmetry and we will show how they allow us to generate just the
Standard Model gauge group. The method involves deforming the known three
generation Calabi Yau model along flat directions. In order to identify
these directions it is necessary to start from a point in moduli space where
the symmetries are larger than for the generic Calabi Yau model because it
is only by finding these symmetries that one may identify the relevant flat
directions. Luckily the superconformal construction \cite{gepner} provides a
way of constructing a 4-d string theory whose large radius limit is just the
3 generation Calabi-Yau theory. Moreover the symmetry of the superconformal
theory is enlarged in just such a way as to allow us to identify absolutely
flat directions in field space. Allowing vevs in these directions still
leaves a viable string theory with three generations but reduces the gauge
symmetry. Such deformed theories need not be continuously connected to the
original theory and provide a new class four dimensional string theories.
Furthermore, as there are a large number of flat directions, there is a rich
diversity of effective low-energy theories related to the original
Calabi-Yau theory. While this may seem disappointing, adding further to the
already vast number of possible string vacua, it does offer some hope for
finding a phenomenologically viable theory. Using this flexibility we show
how it is possible to construct a consistent version of the MSSM in which
the residual low-energy symmetry of the theory guarantees there is just one
pair of light Higgs doublets.

\section{Symmetries and flat directions.\label{sec:flat}}

\subsection{D-flat directions}

We turn now to the central ingredient of the construction of new string
theories, namely the identification of absolutely flat directions in the
scalar potential from the symmetry properties of the theory. For a direction
to be absolutely flat it is necessary for both the F and the D terms to
vanish identically. Let us consider the D terms first. The general form is

\begin{equation}
L_{D}^{Scalar}=\frac{1}{2}\sum_{a}|g_{a}\sum_{i}A_{i}^{\dagger
}T^{a}A_{i}|^{2}
\end{equation}
where $A_{i}$ is the scalar component of the chiral superfield $\phi _{i}$, $%
g_{a}$ is the gauge group coupling constant and $T^{a}$ is the associated
generator. A field direction will be D-flat if the fields acquiring vacuum
expectation values (vevs) along a F-flat direction are singlets under the
gauge group or if there is a cancellation between terms carrying opposite
D-charge. To discuss the latter possibility consider first the simple case
of a single $U(1)$ factor. It is sufficient for D-flatness that there are at
least two fields carrying $U(1)$ charges with opposite signs provided that
the relative magnitude of their vevs are not fixed by the F-terms of the
theory. In this case the ratio of vevs of the two fields is determined by
the condition of D-flatness. This result generalises immediately to more
than one $U(1).$For the non-Abelian case the same conditions apply to the
diagonal generators. Thus one ends up with N constraints amongst the vevs
fields for the case they are charged under $N_{D}$ Abelian gauge group
factors.

\subsection{F-flat directions}

Let us consider now the constraints for F-flatness. From the above
discussion on D-flatness we know that at least one combination of fields
acquiring vevs along D-flat directions will be gauge invariant and therefore
be allowed by the gauge symmetry to appear in the superpotential. For
example, with a single $U(1),$ the condition of D-flatness means there must
be at least two fields, $\phi _{A},\phi _{B}$ carrying opposite sign charges 
$\ Q_{A}$ and -$Q_{B},$ where $Q_{A}$ and $Q_{B}$ are positive. Then the
combination $\phi _{A}^{Q_{B}}\phi _{B}^{Q_{A}}$ is gauge invariant. In
string theory compactifications the charges $Q_{A}$ and $Q_{B}$ are rational
and so the gauge invariant term $\Phi =(\phi _{A}^{Q_{B}}\phi
_{B}^{Q_{A}})^{p}$ for some integer p will involve only integer powers of
the fields and thus be a possible superpotential term. Such a term is not
F-flat by itself because both $F_{\phi _{A}}$ and $F_{\phi _{B}}$ are
non-zero and there is no possibility for other terms to cancel them if only $%
\phi _{A}$ and $\phi _{B}$ have vevs\footnote{%
Apart from the possible cancellation by higher dimension terms giving rise
to a flat ``point'' as discussed at the end of this section}. Of course if $%
\Phi $ has dimension \hbox{$>$}$3$ it will be suppressed by some inverse
power of mass (in general the string scale $\approx $ $M_{Planck})$ but such
a term will spoil the absolute F-flatness we are seeking here. In the case
that additional fields carry vevs there is the possibility of a cancellation
between different contributions to an F-term but this does not solve the
problem because such additional terms necessarily give rise to new F-terms
that are non-vanishing and hence does not lead to F-flatness. For example,
if $\phi _{C}$ has non-vanishing vev and $Q_{C}$ is positive there is an
additional term of the form ($\phi _{C}^{Q_{B}}\phi _{B}^{Q_{C}})^{q}$. \
This contributes to $F_{\phi _{B}}$ and can cancel the contribution from the
first term. However $F_{\phi _{C}}$ is now non zero and so one has not
generated an F-flat potential.

As we have just seen continuous gauge symmetries do not lead to F-flat
directions by themselves. Global symmetries {\it can} evade this conclusion
and ensure F flatness because there is no D-term associated with a global
symmetry and the absence of the D-flatness constraint means that the fields
acquiring vevs can all have the {\it same} sign of global charge. In this
case no combination of charged fields will be neutral and hence invariant
under the global symmetry and thus there is no term in the superpotential
involving these fields. However in string compactification there are no
continuous global symmetries and hence this possibility for
generating flat directions is not realised.The only other symmetries
available to us are discrete symmetries. Non-R symmetries do not lead to
flat directions because we can form an invariant under such a symmetry from
any gauge invariant combination of fields simply by raising it to a suitable
power. For example if the gauge invariant combination of fields, $\Phi ,$
has a non-trivial charge under a $Z_{M}$ symmetry, $\Phi ^{M}$ will be
invariant and a candidate F-term. Thus we are led to the conclusion that the
only symmetry capable of generating a flat direction in the scalar potential
is a discrete R-symmetry \cite{banksdine}. The reason R-symmetries are so
powerful in this respect is easy to see because an R-symmetry requires that
the superspace coordinates, $\theta ,$ transform non-trivially, $\theta
\rightarrow e^{i\beta }\theta $. Thus the superpotential, $P$, must also
transform non-trivially, $P\rightarrow e^{2i\beta }P.$ If the gauge
invariant combination of fields $\Phi $ is invariant under the R-symmetry
then so too is any power of $\Phi ,$ and hence is not a possible F-term. Of
course to ensure F-flatness it is also necessary to avoid any gauge
invariant terms of the form $\phi _{X}\Phi _{i},$ where $\Phi _{i}$ is any
combination of the chiral superfields making up $\Phi $ and $\phi _{X}$ is
any of the chiral superfield which do not acquire vevs. However these latter
terms may be absent due to the gauge and/or discrete symmetries.

\subsection{Moduli}

There is an exception to the conclusion discussed in the last section which
applies to string moduli fields, $m$. These are fields which, due to the
string symmetries, have absolutely flat directions when non-moduli fields
have zero vevs. The reason is because the string symmetries ensure that the
moduli do not appear in the superpotential on their own but only in
non-renormalisable terms in conjunction with non-moduli fields. Because these
fields have absolutely flat potentials their vevs are undetermined in the
absence of \ supersymmetry breaking and they serve to determine the
couplings of the theory. In Calabi Yau compactification there are $h_{2,1}$
complex structure moduli fields whose vevs determine the Yukawa couplings of
the theory. In addition there are $h_{1,1}$ fields whose vevs determine the
shape of the compactification manifold and determine the Kahler structure of
the theory. Since the complex structure moduli fields appear in the
superpotential they may play a role in setting a F-term to zero and thus
should be considered when looking for absolutely flat directions involving
non-moduli fields.

\subsection{Absolutely flat directions}

From these considerations it is straightforward to quantify the conditions
for absolutely flat directions. We start with $N$ fields acquiring vevs. The
condition of D-flatness imposes $N_{D}$ constraints among the vevs. The
condition of F-flatness imposes a further $N_{F}$ constraints corresponding
to $N_{F}$ F-terms which involve (independent) combinations of the $N$
fields only. Clearly the condition for an absolutely flat direction requires

\begin{equation}
N\geq N_{F}+N_{D}  \label{flatcond}
\end{equation}
Equality only occurs if all vevs are determined (strictly this is not a flat
direction but a point in field space) and this necessarily implies one or
more of the non-moduli fields acquires Planck scale vevs. To see this
consider the simplest example of a theory with no R-symmetry with single
chiral superfield, $\phi ,$ which transforms trivially under any of the
symmetries of the theory. The superpotential has the form

\begin{equation}
P=\lambda \phi ^{3}+\widetilde{\lambda }\frac{\phi ^{4}}{M}+...
\label{simple}
\end{equation}
In this case $N=N_{F}=1$ and the vev of $\phi $ is determined. Clearly the
condition that $F_{\phi }$=0 is satisfied only if $\phi $ develops a vev of
O($M$). In this case there is no residual flat direction and the non-trivial
minimum is not connected to the trivial minimum with $\phi =0.$ In this case
identification of a {\it point} at which the vacuum energy vanishes allows
us to build a new string theory not continuously connected to the original.
This is to be contrasted with the case in which the inequality is satisfied
in eq(\ref{flatcond}) may be continuously connected to the original theory.

We wish to explore the possibility that a specific string theory possesses
absolutely flat directions allowing us to construct new string
compactifications\footnote{%
For other work constructing new theories by deforming along flat directions,
see \cite{Brian,cvetic}.}. The first point to notice is that an almost
necessary condition is that there should be no terms allowed in the
superpotential of the form $\Phi ^{n}$ where $\Phi $ is the gauge invariant
combination of fields introduced above. The reason for this is because any
term of this form generates $N_{\Phi }$ F-terms involving the fields
contributing to $\Phi ,$ all of which have non-zero vevs. Thus for this
subset N$_{\Phi }$ of the $N$ fields there are $N_{\Phi }$ F-terms so
by eq(\ref{flatcond}) all the vevs are determined. Thus for this to be an
absolutely flat direction there should be no terms in the superpotential of
the form $\phi _{X}\Phi ^{m}$ for any of the fields $\phi _{X}$ of the
theory. This is a very strong condition and is not realised in a typical
string theory. On the other hand if the term $\Phi ^{n}$ is forbidden then
the condition for a flat direction becomes much simpler, namely the number $%
N_{X}$ of terms of the form $\phi _{X}\Phi ^{m}$ should, by
eq(\ref{flatcond}), be bounded  by the constraint 
\begin{equation}
N_{\phi }\leq N_{X}+N_{D}  \label{flatcond1}
\end{equation}
In Section \ref{sec:3genCY} we will construct explicit examples of a three
generation string theory satisfying these conditions but first we present a
brief review of the three generation string construction that provides the
starting point of our discussion.

\section{The Three generation Calabi Yau Model.\label{sec:model}}

We start with a brief review of the superconformal version of the 3
generation Calabi Yau theory which provides the starting point for our new
(2,0) theories with reduced gauge symmetry. The construction introduced by
Gepner starts with the tensor product of N=1 superconformal theories in 2D.
The resulting theory has an enhanced gauge group $G=E_{8}\otimes
E_{6}\otimes U(1)^{3}.$ In addition it has a much richer class of discrete
symmetries including discrete R-symmetries. The latter is important because,
alone amongst discrete symmetries, it can lead to absolutely flat directions
in field space\footnote{%
Deformations of Calabi-Yau theories along absolutely flat directions ensured
by R-symmetries have previously been considered by Greene \cite{Brian}. This
approach did not start with the superconformal construction used here.}.

The trace anomaly for a level $k$ superconformal theory ($k$ is the
principal quantum number labeling the superconformal theory) is given by 
\begin{equation}
c={\frac{3k}{k+2}}
\end{equation}
By choosing an appropriate set of the minimal $N=2$ superconformal theories,
we are able to provide the $c=9$ central charge. Together with $c=3$
contribution from the free space-time fields, which describe the flat four
dimensional part of the model, the total central charge is 12, which is
required for anomaly cancellation in the heterotic string. In analogy with
heterotic compactifications, we also introduce two sectors (corresponding
respectively to left and right movers) namely the gauge sector and the SUSY
sector. The procedure leading to space-time supersymmetry and the required
gauge group consists of imposing modular invariant constraints on the
quantum numbers of the fields in the theory. This construction has been
described in great details by Gepner \cite{gepner} so here we will only
summarise the results of the analysis.

The primary fields in every minimal superconformal models are labeled by
integers $l$ and $q$. The first one is the principal quantum number, taking
its values in the range $0,...,k$ while the second one labels the $U(1)$%
\footnote{%
The $U(1)$ symmetry is a part of the superconformal algebra associated with
each tensor factor.} charge of the field and is defined modulo $2(k+2)$. In
addition, fields can belong to either Ramond or Neveu-Schwarz sector. We
take this into account by introducing the third label $s$ (0 for
Neveu-Schwarz and $\pm 1$ for Ramond sector). The most important thing in
our construction of new models will be the $U(1)$ charges of the fields.
These are given by 
\begin{equation}
Q=-{\frac{q}{k+2}}+{\frac{s}{2}}.
\end{equation}

The model we consider here is the three generation model, constructed by
Gepner \cite{gepner3g}. This case has been discussed in many papers \cite
{schmidt} and it has been shown that it corresponds to the heterotic string
theory compactified on a specific Calabi-Yau manifold (leading eventually to
the three generations of fermions). The results we quote here are based on
the extensive work of Scheich and Schmidt \cite{scheich}.

The three generation model is constructed by representing the internal
degrees of freedom by a tensor product of four superconformal field
theories, one with level $k=1$ and three with $k=16$. The same product is
used in both supersymmetry and gauge sectors and a modular invariant theory
is formed using the $A$ \cite{gepner} invariant for the $k=1$ factor and $%
E_{7}$ affine modular invariants for $k=16$ factors. At this stage the model
contains 35 generations (${\bf 27}$ of $E_{6}$), 8 anti-generations ($%
\overline{{\bf 27}}${\bf \ }of $E_{6}$) and 197 massless $E_{6}$ singlets.
To reduce the net number of generation to three we mod out the tensor
product of the superconformal theories by two discrete symmetries present in
the model $S$, which are the cyclic permutations of the three $k=16$
sub-theories and a $Z_{3}$ symmetry generated by (0,3,6,0) element (the four
numbers correspond respectively to $Z_{3}$ charge in the first subtheory and
the three $Z_{18}$ charges in the remaining ones) \cite{gepner}.

The fields are labeled by the quantum numbers of the superconformal factors
in the form 
\begin{equation}
\left( 
\begin{array}{ccc}
l & q & s \\ 
{\overline{l}} & {\overline{q}} & {\overline{s}}
\end{array}
\right) .  \label{state}
\end{equation}
The notation used here refers to the $SO(10)$ {\bf 10 }component in the
decomposition of the full {\bf 27}-plet of $E_{6}$ into the subgroup $%
SO(10)\times U(1)$. To get the full 27-plet, one has to take three different
fields corresponding respectively to the {\bf 1}, {\bf 10} and {\bf 16} of $%
SO(10)$. These three different representations can be generated one from
each other by acting with an operator which results in shifting ${\overline{q%
}}\rightarrow {\overline{q}}+1$ and ${\overline{s}}\rightarrow {\overline{s}}%
+1$. Applying this successively generates the set {\bf 10}$\rightarrow ${\bf %
16}$\rightarrow ${\bf 1}$\rightarrow ${\bf 10}.

An analogous procedure applies to the supersymmetry sector. A given field,
described by $({\overline{l}},{\overline{q}},{\overline{s}})$ corresponds to
only one (for example scalar) component of the chiral superfield. To get the
full supermultiplet, one has to sum three different Lorentz states. Again,
we obtain them using a specific operator (in this case it is the
supersymmetry charge), which shifts the quantum number in the SUSY sector: $%
q\rightarrow q+1$, $s\rightarrow s+1$. The notation of eq(\ref{state})
refers to the scalar component. From it we generate the one corresponding to
fermions (in the $-{\frac{1}{2}}$ picture) by applying the supersymmetry
charge operator once. Applying the supersymmetry charge twice yields the
vertex operator for the auxiliary F-field.

For the scalar components of the chiral supermultiplets which transform
under $SO(10)$ as the {\bf 10} component of an $E_{6}$ {\bf 27}
supermultiplet the massless fields are given in Table \ref{Table:1}

\begin{table}[tbp] \centering%
%
\begin{eqnarray*}
&&l_{1}:\left( 
\begin{array}{ccc}
1 & 1 & 0 \\ 
1 & 1 & 0
\end{array}
\right) \left( 
\begin{array}{ccc}
0 & 0 & 0 \\ 
0 & 0 & 0
\end{array}
\right) \left( 
\begin{array}{ccc}
0 & 0 & 0 \\ 
0 & 0 & 0
\end{array}
\right) \left( 
\begin{array}{ccc}
12 & 12 & 0 \\ 
12 & 12 & 0
\end{array}
\right) \\
&&l_{2}:\left( 
\begin{array}{ccc}
1 & 1 & 0 \\ 
1 & 1 & 0
\end{array}
\right) \left( 
\begin{array}{ccc}
0 & 0 & 0 \\ 
0 & 0 & 0
\end{array}
\right) \left( 
\begin{array}{ccc}
6 & 6 & 0 \\ 
6 & 6 & 0
\end{array}
\right) \left( 
\begin{array}{ccc}
6 & 6 & 0 \\ 
6 & 6 & 0
\end{array}
\right) \\
&&l_{3}:\left( 
\begin{array}{ccc}
1 & 1 & 0 \\ 
1 & 1 & 0
\end{array}
\right) \left( 
\begin{array}{ccc}
4 & 4 & 0 \\ 
4 & 4 & 0
\end{array}
\right) \left( 
\begin{array}{ccc}
4 & 4 & 0 \\ 
4 & 4 & 0
\end{array}
\right) \left( 
\begin{array}{ccc}
4 & 4 & 0 \\ 
4 & 4 & 0
\end{array}
\right) \\
&&l_{4}:\left( 
\begin{array}{ccc}
0 & 0 & 0 \\ 
0 & 0 & 0
\end{array}
\right) \left( 
\begin{array}{ccc}
0 & 0 & 0 \\ 
0 & 0 & 0
\end{array}
\right) \left( 
\begin{array}{ccc}
6 & 6 & 0 \\ 
6 & 6 & 0
\end{array}
\right) \left( 
\begin{array}{ccc}
12 & 12 & 0 \\ 
12 & 12 & 0
\end{array}
\right) \\
&&l_{5}:\left( 
\begin{array}{ccc}
0 & 0 & 0 \\ 
0 & 0 & 0
\end{array}
\right) \left( 
\begin{array}{ccc}
0 & 0 & 0 \\ 
0 & 0 & 0
\end{array}
\right) \left( 
\begin{array}{ccc}
12 & 12 & 0 \\ 
12 & 12 & 0
\end{array}
\right) \left( 
\begin{array}{ccc}
6 & 6 & 0 \\ 
6 & 6 & 0
\end{array}
\right) \\
&&l_{6}:\left( 
\begin{array}{ccc}
0 & 0 & 0 \\ 
0 & 0 & 0
\end{array}
\right) \left( 
\begin{array}{ccc}
4 & 4 & 0 \\ 
4 & 4 & 0
\end{array}
\right) \left( 
\begin{array}{ccc}
4 & 4 & 0 \\ 
4 & 4 & 0
\end{array}
\right) \left( 
\begin{array}{ccc}
10 & 10 & 0 \\ 
10 & 10 & 0
\end{array}
\right) \\
&&l_{7}:\left( 
\begin{array}{ccc}
0 & 0 & 0 \\ 
0 & 0 & 0
\end{array}
\right) \left( 
\begin{array}{ccc}
6 & 6 & 0 \\ 
6 & 6 & 0
\end{array}
\right) \left( 
\begin{array}{ccc}
6 & 6 & 0 \\ 
6 & 6 & 0
\end{array}
\right) \left( 
\begin{array}{ccc}
6 & 6 & 0 \\ 
6 & 6 & 0
\end{array}
\right) \\
&&l_{8}:\left( 
\begin{array}{ccc}
0 & 0 & 0 \\ 
0 & 0 & 0
\end{array}
\right) \left( 
\begin{array}{ccc}
2 & 2 & 0 \\ 
8 & 8 & 0
\end{array}
\right) \left( 
\begin{array}{ccc}
8 & 8 & 0 \\ 
8 & 8 & 0
\end{array}
\right) \left( 
\begin{array}{ccc}
8 & 8 & 0 \\ 
2 & 2 & 0
\end{array}
\right) \\
&&l_{9}:\left( 
\begin{array}{ccc}
0 & 0 & 0 \\ 
0 & 0 & 0
\end{array}
\right) \left( 
\begin{array}{ccc}
2 & 2 & 0 \\ 
8 & 8 & 0
\end{array}
\right) \left( 
\begin{array}{ccc}
8 & 8 & 0 \\ 
2 & 2 & 0
\end{array}
\right) \left( 
\begin{array}{ccc}
8 & 8 & 0 \\ 
8 & 8 & 0
\end{array}
\right) ,
\end{eqnarray*}
\caption{ The massless  fields of the three generation Gepner construction transforming as the $27$ representation of $E_6$. The four factors refer to the $1.16^3$ superconformal factors of the construction.
The quantum numbers given refer to the scalar component transforming as the 10 of the SO(10) subgroup of $E_6$. \label{Table:1}}%
\end{table}%
%

It is important to remember that in all these fields (and the ones which are
listed further below), the three k=16 factors can be interchanged. This may
give one, three or six different fields depending on whether the the quantum
numbers in all the factors are the same, two are the same or all are
different respectively. Upon modding out by the cyclic permutations, the
surviving fields are the symmetric combinations. If Wilson lines are
associated with the cyclic permutations then the states left light will
correspond to different combinations such that the product of the discrete
gauge group factor and the permutation are singlet.

The {\bf 27} mirror generations are given in Table \ref{Table:2}.
\begin{table}[tbp] \centering%
%
\begin{eqnarray*}
&&{\overline{l}}_{1}:\left( 
\begin{array}{ccc}
0 & 0 & 0 \\ 
0 & 2 & 2
\end{array}
\right) \left( 
\begin{array}{ccc}
2 & 2 & 0 \\ 
8 & -8 & 0
\end{array}
\right) \left( 
\begin{array}{ccc}
8 & 8 & 0 \\ 
2 & -2 & 0
\end{array}
\right) \left( 
\begin{array}{ccc}
8 & 8 & 0 \\ 
2 & -2 & 0
\end{array}
\right) \\
&&{\overline{l}}_{2}:\left( 
\begin{array}{ccc}
1 & 1 & 0 \\ 
1 & -1 & 0
\end{array}
\right) \left( 
\begin{array}{ccc}
4 & 4 & 0 \\ 
4 & -4 & 0
\end{array}
\right) \left( 
\begin{array}{ccc}
4 & 4 & 0 \\ 
4 & -4 & 0
\end{array}
\right) \left( 
\begin{array}{ccc}
4 & 4 & 0 \\ 
4 & -4 & 0
\end{array}
\right) \\
&&{\overline{l}}_{3}:\left( 
\begin{array}{ccc}
1 & 1 & 0 \\ 
1 & 3 & 2
\end{array}
\right) \left( 
\begin{array}{ccc}
2 & 2 & 0 \\ 
8 & -8 & 0
\end{array}
\right) \left( 
\begin{array}{ccc}
2 & 2 & 0 \\ 
8 & -8 & 0
\end{array}
\right) \left( 
\begin{array}{ccc}
8 & 8 & 0 \\ 
2 & -2 & 0
\end{array}
\right) \\
&&{\overline{l}}_{4}:\left( 
\begin{array}{ccc}
0 & 0 & 0 \\ 
0 & 0 & 0
\end{array}
\right) \left( 
\begin{array}{ccc}
6 & 6 & 0 \\ 
6 & -6 & 0
\end{array}
\right) \left( 
\begin{array}{ccc}
6 & 6 & 0 \\ 
6 & -6 & 0
\end{array}
\right) \left( 
\begin{array}{ccc}
6 & 6 & 0 \\ 
6 & -6 & 0
\end{array}
\right) \\
&&{\overline{l}}_{5}:\left( 
\begin{array}{ccc}
0 & 0 & 0 \\ 
0 & 0 & 0
\end{array}
\right) \left( 
\begin{array}{ccc}
0 & 0 & 0 \\ 
0 & 0 & 0
\end{array}
\right) \left( 
\begin{array}{ccc}
6 & 6 & 0 \\ 
6 & -6 & 0
\end{array}
\right) \left( 
\begin{array}{ccc}
12 & 12 & 0 \\ 
12 & -12 & 0
\end{array}
\right) \\
&&{\overline{l}}_{6}:\left( 
\begin{array}{ccc}
0 & 0 & 0 \\ 
0 & 0 & 0
\end{array}
\right) \left( 
\begin{array}{ccc}
0 & 0 & 0 \\ 
0 & 0 & 0
\end{array}
\right) \left( 
\begin{array}{ccc}
12 & 12 & 0 \\ 
12 & -12 & 0
\end{array}
\right) \left( 
\begin{array}{ccc}
6 & 6 & 0 \\ 
6 & -6 & 0
\end{array}
\right)
\end{eqnarray*}
\caption{The massless  fields of the three generation Gepner construction transforming as the $\bar{27}$ representations of $E_6$. The four factors refer to the $1.16^3$ superconformal factors of the construction.
The quantum numbers given refer to the scalar component transforming as the 10 of the SO(10) subgroup of $E_6$\label{Table:2}}%
\end{table}%
%
Finally the gauge singlets are given in Table \ref{Table:singlets}

\begin{table}[tbp] \centering%
%
\begin{eqnarray*}
&&\phi _{1}:\left( 
\begin{array}{ccc}
0 & 0 & 0 \\ 
0 & 0 & 0
\end{array}
\right) \left( 
\begin{array}{ccc}
0 & 0 & 0 \\ 
0 & 0 & 0
\end{array}
\right) \left( 
\begin{array}{ccc}
6 & 6 & 0 \\ 
6 & 0 & 0
\end{array}
\right) \left( 
\begin{array}{ccc}
12 & 12 & 0 \\ 
4 & 0 & 0
\end{array}
\right) \\
&&\phi _{2}:\left( 
\begin{array}{ccc}
0 & 0 & 0 \\ 
0 & 0 & 0
\end{array}
\right) \left( 
\begin{array}{ccc}
0 & 0 & 0 \\ 
0 & 0 & 0
\end{array}
\right) \left( 
\begin{array}{ccc}
12 & 12 & 0 \\ 
4 & 0 & 0
\end{array}
\right) \left( 
\begin{array}{ccc}
6 & 6 & 0 \\ 
6 & 0 & 0
\end{array}
\right) \\
&&\phi _{3}:\left( 
\begin{array}{ccc}
0 & 0 & 0 \\ 
0 & 0 & 0
\end{array}
\right) \left( 
\begin{array}{ccc}
2 & 2 & 0 \\ 
8 & -4 & 0
\end{array}
\right) \left( 
\begin{array}{ccc}
8 & 8 & 0 \\ 
2 & 2 & 0
\end{array}
\right) \left( 
\begin{array}{ccc}
8 & 8 & 0 \\ 
2 & 2 & 0
\end{array}
\right) \\
&&\phi _{4}:\left( 
\begin{array}{ccc}
0 & 0 & 0 \\ 
0 & 0 & 0
\end{array}
\right) \left( 
\begin{array}{ccc}
4 & 4 & 0 \\ 
4 & -2 & 0
\end{array}
\right) \left( 
\begin{array}{ccc}
4 & 4 & 0 \\ 
4 & -2 & 0
\end{array}
\right) \left( 
\begin{array}{ccc}
10 & 10 & 0 \\ 
6 & 4 & 0
\end{array}
\right) \\
&&\phi _{5}:\left( 
\begin{array}{ccc}
0 & 0 & 0 \\ 
0 & 0 & 0
\end{array}
\right) \left( 
\begin{array}{ccc}
6 & 6 & 0 \\ 
6 & 6 & 0
\end{array}
\right) \left( 
\begin{array}{ccc}
6 & 6 & 0 \\ 
6 & -6 & 0
\end{array}
\right) \left( 
\begin{array}{ccc}
6 & 6 & 0 \\ 
6 & 0 & 0
\end{array}
\right) \\
&&\phi _{6}:\left( 
\begin{array}{ccc}
0 & 0 & 0 \\ 
0 & 0 & 0
\end{array}
\right) \left( 
\begin{array}{ccc}
6 & 6 & 0 \\ 
6 & -6 & 0
\end{array}
\right) \left( 
\begin{array}{ccc}
6 & 6 & 0 \\ 
6 & 6 & 0
\end{array}
\right) \left( 
\begin{array}{ccc}
6 & 6 & 0 \\ 
6 & 0 & 0
\end{array}
\right) \\
&&\phi _{7}:\left( 
\begin{array}{ccc}
0 & 0 & 0 \\ 
0 & 2 & 2
\end{array}
\right) \left( 
\begin{array}{ccc}
4 & 4 & 0 \\ 
4 & 0 & 0
\end{array}
\right) \left( 
\begin{array}{ccc}
4 & 4 & 0 \\ 
4 & 0 & 0
\end{array}
\right) \left( 
\begin{array}{ccc}
10 & 10 & 0 \\ 
6 & 6 & 0
\end{array}
\right) \\
&&\phi _{8}:\left( 
\begin{array}{ccc}
0 & 0 & 0 \\ 
0 & 2 & 2
\end{array}
\right) \left( 
\begin{array}{ccc}
4 & 4 & 0 \\ 
4 & 0 & 0
\end{array}
\right) \left( 
\begin{array}{ccc}
4 & 4 & 0 \\ 
12 & 12 & 0
\end{array}
\right) \left( 
\begin{array}{ccc}
10 & 10 & 0 \\ 
6 & -6 & 0
\end{array}
\right) \\
&&\phi _{9}:\left( 
\begin{array}{ccc}
0 & 0 & 0 \\ 
0 & 2 & 2
\end{array}
\right) \left( 
\begin{array}{ccc}
4 & 4 & 0 \\ 
12 & 12 & 0
\end{array}
\right) \left( 
\begin{array}{ccc}
4 & 4 & 0 \\ 
4 & 0 & 0
\end{array}
\right) \left( 
\begin{array}{ccc}
10 & 10 & 0 \\ 
6 & -6 & 0
\end{array}
\right) \\
&&\phi _{10}:\left( 
\begin{array}{ccc}
0 & 0 & 0 \\ 
0 & 2 & 2
\end{array}
\right) \left( 
\begin{array}{ccc}
4 & 4 & 0 \\ 
12 & -12 & 0
\end{array}
\right) \left( 
\begin{array}{ccc}
4 & 4 & 0 \\ 
12 & 12 & 0
\end{array}
\right) \left( 
\begin{array}{ccc}
10 & 10 & 0 \\ 
6 & 6 & 0
\end{array}
\right) \\
&&\phi _{11}:\left( 
\begin{array}{ccc}
0 & 0 & 0 \\ 
0 & 2 & 2
\end{array}
\right) \left( 
\begin{array}{ccc}
4 & 4 & 0 \\ 
12 & 12 & 0
\end{array}
\right) \left( 
\begin{array}{ccc}
4 & 4 & 0 \\ 
12 & -12 & 0
\end{array}
\right) \left( 
\begin{array}{ccc}
10 & 10 & 0 \\ 
6 & 6 & 0
\end{array}
\right) \\
&&\phi _{12}:\left( 
\begin{array}{ccc}
0 & 0 & 0 \\ 
0 & 4 & 2
\end{array}
\right) \left( 
\begin{array}{ccc}
2 & 2 & 0 \\ 
8 & -6 & 0
\end{array}
\right) \left( 
\begin{array}{ccc}
8 & 8 & 0 \\ 
2 & 0 & 0
\end{array}
\right) \left( 
\begin{array}{ccc}
8 & 8 & 0 \\ 
2 & 0 & 0
\end{array}
\right) \\
&&\phi _{13}:\left( 
\begin{array}{ccc}
0 & 0 & 0 \\ 
0 & 4 & 2
\end{array}
\right) \left( 
\begin{array}{ccc}
4 & 4 & 0 \\ 
4 & -4 & 0
\end{array}
\right) \left( 
\begin{array}{ccc}
4 & 4 & 0 \\ 
4 & 2 & 0
\end{array}
\right) \left( 
\begin{array}{ccc}
10 & 10 & 0 \\ 
6 & -4 & 0
\end{array}
\right) \\
&&\phi _{14}:\left( 
\begin{array}{ccc}
0 & 0 & 0 \\ 
0 & 4 & 2
\end{array}
\right) \left( 
\begin{array}{ccc}
4 & 4 & 0 \\ 
4 & 2 & 0
\end{array}
\right) \left( 
\begin{array}{ccc}
4 & 4 & 0 \\ 
4 & -4 & 0
\end{array}
\right) \left( 
\begin{array}{ccc}
10 & 10 & 0 \\ 
6 & -4 & 0
\end{array}
\right) \\
&&\phi _{15}:\left( 
\begin{array}{ccc}
1 & 1 & 0 \\ 
1 & 3 & 2
\end{array}
\right) \left( 
\begin{array}{ccc}
4 & 4 & 0 \\ 
4 & 0 & 0
\end{array}
\right) \left( 
\begin{array}{ccc}
4 & 4 & 0 \\ 
4 & 0 & 0
\end{array}
\right) \left( 
\begin{array}{ccc}
4 & 4 & 0 \\ 
4 & 0 & 0
\end{array}
\right) \\
&&\phi _{16}:\left( 
\begin{array}{ccc}
1 & 1 & 0 \\ 
1 & 3 & 2
\end{array}
\right) \left( 
\begin{array}{ccc}
4 & 4 & 0 \\ 
4 & 0 & 0
\end{array}
\right) \left( 
\begin{array}{ccc}
4 & 4 & 0 \\ 
12 & -12 & 0
\end{array}
\right) \left( 
\begin{array}{ccc}
4 & 4 & 0 \\ 
12 & 12 & 0
\end{array}
\right) \\
&&\phi _{17}:\left( 
\begin{array}{ccc}
1 & 1 & 0 \\ 
1 & 3 & 2
\end{array}
\right) \left( 
\begin{array}{ccc}
4 & 4 & 0 \\ 
12 & -12 & 0
\end{array}
\right) \left( 
\begin{array}{ccc}
4 & 4 & 0 \\ 
4 & 0 & 0
\end{array}
\right) \left( 
\begin{array}{ccc}
4 & 4 & 0 \\ 
12 & 12 & 0
\end{array}
\right) \\
&&\phi _{18}:\left( 
\begin{array}{ccc}
1 & 1 & 0 \\ 
1 & 1 & 0
\end{array}
\right) \left( 
\begin{array}{ccc}
0 & 0 & 0 \\ 
0 & 0 & 0
\end{array}
\right) \left( 
\begin{array}{ccc}
6 & 6 & 0 \\ 
6 & 0 & 0
\end{array}
\right) \left( 
\begin{array}{ccc}
6 & 6 & 0 \\ 
6 & -6 & 0
\end{array}
\right) \\
\end{eqnarray*}
\end{table}%
%

\bigskip 
\begin{table}[tbp] \centering%
%
\begin{eqnarray*}
&&\phi _{19}:\left( 
\begin{array}{ccc}
1 & 1 & 0 \\ 
1 & 1 & 0
\end{array}
\right) \left( 
\begin{array}{ccc}
0 & 0 & 0 \\ 
0 & 0 & 0
\end{array}
\right) \left( 
\begin{array}{ccc}
6 & 6 & 0 \\ 
6 & -6 & 0
\end{array}
\right) \left( 
\begin{array}{ccc}
6 & 6 & 0 \\ 
6 & 0 & 0
\end{array}
\right) \\
&&\phi _{20}:\left( 
\begin{array}{ccc}
1 & 1 & 0 \\ 
1 & 1 & 0
\end{array}
\right) \left( 
\begin{array}{ccc}
4 & 4 & 0 \\ 
4 & -2 & 0
\end{array}
\right) \left( 
\begin{array}{ccc}
4 & 4 & 0 \\ 
4 & -2 & 0
\end{array}
\right) \left( 
\begin{array}{ccc}
4 & 4 & 0 \\ 
4 & -2 & 0
\end{array}
\right) \\
&&\phi _{21}:\left( 
\begin{array}{ccc}
0 & 0 & 0 \\ 
0 & 4 & 2
\end{array}
\right) \left( 
\begin{array}{ccc}
4 & 4 & 0 \\ 
4 & 2 & 0
\end{array}
\right) \left( 
\begin{array}{ccc}
4 & 4 & 0 \\ 
4 & 2 & 0
\end{array}
\right) \left( 
\begin{array}{ccc}
10 & 10 & 0 \\ 
10 & -10 & 0
\end{array}
\right) \\
&&\phi _{22}:\left( 
\begin{array}{ccc}
0 & 0 & 0 \\ 
0 & 4 & 2
\end{array}
\right) \left( 
\begin{array}{ccc}
4 & 4 & 0 \\ 
4 & -4 & 0
\end{array}
\right) \left( 
\begin{array}{ccc}
4 & 4 & 0 \\ 
4 & -4 & 0
\end{array}
\right) \left( 
\begin{array}{ccc}
10 & 10 & 0 \\ 
6 & 2 & 0
\end{array}
\right) \\
&&\phi _{23}:\left( 
\begin{array}{ccc}
0 & 0 & 0 \\ 
0 & 0 & 0
\end{array}
\right) \left( 
\begin{array}{ccc}
4 & 4 & 0 \\ 
4 & 4 & 0
\end{array}
\right) \left( 
\begin{array}{ccc}
4 & 4 & 0 \\ 
4 & 4 & 0
\end{array}
\right) \left( 
\begin{array}{ccc}
10 & 10 & 0 \\ 
10 & -8 & 0
\end{array}
\right) \\
&&\phi _{24}:\left( 
\begin{array}{ccc}
1 & 1 & 0 \\ 
1 & -1 & 0
\end{array}
\right) \left( 
\begin{array}{ccc}
4 & 4 & 0 \\ 
4 & 2 & 0
\end{array}
\right) \left( 
\begin{array}{ccc}
4 & 4 & 0 \\ 
4 & 2 & 0
\end{array}
\right) \left( 
\begin{array}{ccc}
4 & 4 & 0 \\ 
4 & 2 & 0
\end{array}
\right) \\
&&\phi _{25}:\left( 
\begin{array}{ccc}
0 & 0 & 0 \\ 
0 & 0 & 0
\end{array}
\right) \left( 
\begin{array}{ccc}
4 & 4 & 0 \\ 
4 & 4 & 0
\end{array}
\right) \left( 
\begin{array}{ccc}
10 & 10 & 0 \\ 
6 & -2 & 0
\end{array}
\right) \left( 
\begin{array}{ccc}
4 & 4 & 0 \\ 
4 & -2 & 0
\end{array}
\right) \\
&&\phi _{26}:\left( 
\begin{array}{ccc}
0 & 0 & 0 \\ 
0 & 0 & 0
\end{array}
\right) \left( 
\begin{array}{ccc}
4 & 4 & 0 \\ 
4 & 4 & 0
\end{array}
\right) \left( 
\begin{array}{ccc}
4 & 4 & 0 \\ 
4 & -2 & 0
\end{array}
\right) \left( 
\begin{array}{ccc}
10 & 10 & 0 \\ 
6 & -2 & 0
\end{array}
\right)
\end{eqnarray*}
\caption{The scalar components of the massless $E_6$ gauge singlet
fields.\label{Table:singlets}}%
\end{table}%
%

In addition there are 9 complex structure moduli fields which determine the
Yukawa couplings of the theory and 6 Kahler structure moduli which determine
the metric.

In the construction of phenomenologically relevant models, one has to break
the gauge group from $E_{6}$ to a smaller group leading, in the end, to the
Standard Model gauge group. In the case of Calabi-Yau compactifications, the
best way to do it \cite{rosspok} is to use Wilson lines. In the Gepner model
there is an exact analog of this mechanism \cite{alwis} which, as in the
Calabi-Yau case, requires the introduction of additional fields with
non-trivial transformation properties under the action of the discrete group
which we embed in the gauge group. In the case this is the $Z_{3}$ cyclic
permutation group, the new fields carry the quantum numbers of the
left-handed quarks, $q$, and the left-handed antiquarks, $Q$. They are given
by the combinations of the fields given above which transform as $\alpha $
and $\alpha ^{2}$ ($\alpha =e^{i2\pi /3}$) respectively under the cyclic
permutation group.

For completeness we list in Table \ref{Table:4} the fields appearing in the
case that the discrete group associated with the Wilson lines is the phase
twist generated by (0,3,6,0) element \cite{scheich}.

\bigskip 
\begin{table}[tbp] \centering%
%
\begin{eqnarray*}
&&q_{1}:\left( 
\begin{array}{ccc}
1 & 1 & 0 \\ 
1 & 1 & 0
\end{array}
\right) \left( 
\begin{array}{ccc}
8 & 8 & 0 \\ 
8 & 8 & 0
\end{array}
\right) \left( 
\begin{array}{ccc}
4 & 4 & 0 \\ 
4 & 4 & 0
\end{array}
\right) \left( 
\begin{array}{ccc}
0 & 0 & 0 \\ 
0 & 0 & 0
\end{array}
\right) \\
&&q_{2}:\left( 
\begin{array}{ccc}
0 & 0 & 0 \\ 
0 & 0 & 0
\end{array}
\right) \left( 
\begin{array}{ccc}
0 & 0 & 0 \\ 
0 & 0 & 0
\end{array}
\right) \left( 
\begin{array}{ccc}
8 & 8 & 0 \\ 
8 & 8 & 0
\end{array}
\right) \left( 
\begin{array}{ccc}
10 & 10 & 0 \\ 
10 & 10 & 0
\end{array}
\right) \\
&&q_{3}:\left( 
\begin{array}{ccc}
0 & 0 & 0 \\ 
0 & 0 & 0
\end{array}
\right) \left( 
\begin{array}{ccc}
4 & 4 & 0 \\ 
4 & 4 & 0
\end{array}
\right) \left( 
\begin{array}{ccc}
6 & 6 & 0 \\ 
6 & 6 & 0
\end{array}
\right) \left( 
\begin{array}{ccc}
8 & 8 & 0 \\ 
8 & 8 & 0
\end{array}
\right)
\end{eqnarray*}
\begin{eqnarray*}
&&Q_{1}:\left( 
\begin{array}{ccc}
1 & 1 & 0 \\ 
1 & 1 & 0
\end{array}
\right) \left( 
\begin{array}{ccc}
0 & 0 & 0 \\ 
0 & 0 & 0
\end{array}
\right) \left( 
\begin{array}{ccc}
4 & 4 & 0 \\ 
4 & 4 & 0
\end{array}
\right) \left( 
\begin{array}{ccc}
8 & 8 & 0 \\ 
8 & 8 & 0
\end{array}
\right) \\
&&Q_{2}:\left( 
\begin{array}{ccc}
0 & 0 & 0 \\ 
0 & 0 & 0
\end{array}
\right) \left( 
\begin{array}{ccc}
10 & 10 & 0 \\ 
10 & 10 & 0
\end{array}
\right) \left( 
\begin{array}{ccc}
8 & 8 & 0 \\ 
8 & 8 & 0
\end{array}
\right) \left( 
\begin{array}{ccc}
0 & 0 & 0 \\ 
0 & 0 & 0
\end{array}
\right) \\
&&Q_{3}:\left( 
\begin{array}{ccc}
0 & 0 & 0 \\ 
0 & 0 & 0
\end{array}
\right) \left( 
\begin{array}{ccc}
8 & 8 & 0 \\ 
8 & 8 & 0
\end{array}
\right) \left( 
\begin{array}{ccc}
6 & 6 & 0 \\ 
6 & 6 & 0
\end{array}
\right) \left( 
\begin{array}{ccc}
4 & 4 & 0 \\ 
4 & 4 & 0
\end{array}
\right) .
\end{eqnarray*}
\caption{Quark and lepton fields for the case of Wilson line breaking with the twist generated by the (0,3,6,0) element.
\label{Table:4}}%
\end{table}%
%

\subsection{Couplings in the Gepner model.}

The crucial element in our determination of flat directions in the Gepner
model is the identification of the discrete symmetries of the model. Thus we
wish to determine the correlation functions of the form 
\begin{equation}
<V_{-1}^{1}V_{-1/2}^{2}V_{-1/2}^{3}V_{0}^{4}...V_{0}^{n}>,  \label{corrfn}
\end{equation}
where $V_{-1(0)}$ is a vertex operator for a space-time scalar in the $%
(-1)((0))$ picture and $V_{-1/2}$ is a vertex operator for a space-time
fermion in the $(-1/2)$ picture. The choice of vertex operators in the (0)
picture for $n>3$ is dictated by the need to have the correct ghost charge
(-2) for the non-vanishing correlator computed at the string tree level \cite
{friedman}. 
All the quantum numbers listed above corresponded to vertex operators for
the space-time scalars in the (-1) picture 
\begin{equation}
V_{-1}\sim \bigotimes_{i=1}^{4}\left( 
\begin{array}{ccc}
l_{i} & q_{i} & s_{i} \\ 
{\overline{l}}_{i} & {\overline{q}}_{i} & {\overline{s}}_{i}
\end{array}
\right) .
\end{equation}
As mentioned above, to obtain vertex operators for fermions in the $(-{\frac{%
1}{2}})$ picture we have to act with the supersymmetry charge which shifts
the quantum numbers giving 
\begin{equation}
V_{-1/2}\sim \bigotimes_{i=1}^{4}\left( 
\begin{array}{ccc}
l_{i} & q_{i}+1 & s_{i}+1 \\ 
{\overline{l}}_{i} & {\overline{q}}_{i} & {\overline{s}}_{i}
\end{array}
\right) .
\end{equation}
The transformation from the (-1) picture to the (0) picture is carried out
by one of the superpartners of the stress tensor \cite{distler} and results
in the fields with the following quantum numbers 
\begin{equation}
V_{0}\sim \sum_{j=1}^{4}\bigotimes_{i=1}^{j-1}\left( 
\begin{array}{ccc}
l_{i} & q_{i} & s_{i} \\ 
{\overline{l}}_{i} & {\overline{q}}_{i} & {\overline{s}}_{i}
\end{array}
\right) \otimes \left( 
\begin{array}{ccc}
l_{j} & q_{j} & s_{j}+2 \\ 
{\overline{l}}_{j} & {\overline{q}}_{j} & {\overline{s}}_{j}
\end{array}
\right) \otimes \bigotimes_{i=j+1}^{4}\left( 
\begin{array}{ccc}
l_{i} & q_{i} & s_{i} \\ 
{\overline{l}}_{i} & {\overline{q}}_{i} & {\overline{s}}_{i}
\end{array}
\right) .
\end{equation}

As discussed in detail in \cite{lutross}, the allowed correlation functions
of eq(\ref{corrfn}) are entirely specified by the symmetries present in the
model. Of particular importance are the $U(1)$ factors associated with each $%
N=2$ superconformal factor.In order to discuss these conditions it is
convenient to define the $U(1)$ charges for all the subtheories 
\begin{eqnarray}
\alpha _{i}^{J} &=&-{\frac{q_{i}^{J}}{k_{i}+2}}+{\frac{s_{i}^{J}}{2}}
\label{alpha} \\
\overline{\alpha }_{i}^{J} &=&-{\frac{\overline{q}_{i}^{J}}{k_{i}+2}}+{\frac{%
\overline{s}_{i}^{J}}{2}}  \label{alphab}
\end{eqnarray}
The index $i$ labels the subtheories, while $J$ refers to the different
fields entering the coupling.

In the case of \ the SUSY sector, the condition is the same for all types of
couplings and reflects the picture changing needed to provide the total
ghost charge equal -2. Taking into account the rules of picture changing
mentioned earlier in this section, we obtain the following conservation law 
\begin{equation}
\sum_{J=1}^{n}\left( \alpha _{i}^{J}-{\frac{2}{k_{i}+2}}+1\right)
+\sum_{J=4}^{n}d_{i}^{J}=0\;\;\;\;(\forall i).  \label{rule1}
\end{equation}
In the last term, $d_{i}^{J}$ can be equal 0 or 1 with the condition 
\begin{equation}
\sum_{i=1}^{4}\sum_{J=1}^{n}d_{i}^{J}=n-3.  \label{rule2}
\end{equation}
Obviously, this condition exists only for $n>3$ (it is only when we have
more then three fields in the coupling that we have to change the picture to
(0)).

In the case of the gauge sector, the conservation rule depends on the type
of coupling we are discussing. As we mentioned above, the quantum numbers of
the fields listed in the Tables correspond to the {\bf 10} representation of
the $SO(10)$ factor of $E_{6}$. This means that for example in the case of
the coupling ${\bf 27\times 27\times 27}$, in order to make a gauge
invariant coupling we have to change two of the three fields from {\bf 10} to 
{\bf 16}. This gives us 
\begin{equation}
\sum_{J=1}^{n}\left( {\overline{\alpha }}_{i}^{J}-{\frac{2}{k_{i}+2}}%
+1\right) =0\;\;\;\;(\forall i).  \label{rule3}
\end{equation}
In the case of couplings of the type ${\bf 1 \times 27\times }\overline{%
{\bf 27}}$ there is no need to change any representations because the
quantum numbers for the fields listed in the Tables generate the term{\bf \ }%
${\bf 1\times 10\times }\overline{{\bf 10}}${\ which is already gauge
invariant. This leads to the simple conservation law 
\begin{equation}
\sum_{J=1}^{n}{\overline{\alpha }}_{i}^{J}=0\;\;\;\;(\forall i).
\label{rule4}
\end{equation}
The same is obviously true for the ${\bf 1\times 1\times 1}${\bf \ }%
couplings. }

As we shall now discuss the constraints of eqs(\ref{rule1},
\ref{rule2}, \ref{rule3}, \ref{rule4}) are quite restrictive and lead
to absolutely flat
directions of the type discussed in Section \ref{sec:flat}.

\subsection{Flat directions\label{sec:flatex}}

We start this section with a simple example of a flat direction involving $%
E_{6}$ non-singlet fields. In particular we consider the case that the $N$
component of the fields $l_{7}$ and ${\overline{l}}_{4}$ acquires a non-zero
vev.

We first establish that there are no terms allowed in the superpotential of
the form $(l_{7}\overline{l}_{4})^{m}.$ From Table \ref{Table:1} and \ref
{Table:2} we see that these fields both have vanishing $\alpha _{1}$ charge.
>From eq(\ref{rule1}) we see that for a non-vanishing correlator the sum of
the $\alpha _{1}$ charges should be $\frac{1}{3},$ $Mod$ $\ 3$, i.e. the
symmetry is an R-symmetry. Thus we see immediately that no terms are allowed
of the form $(l_{7}\overline{l}_{4})^{m}.$ This establishes $l_{7},$ $%
\overline{l}_{4}$ as a good candidate for an absolutely flat direction.
However to complete the proof that this is the case we need to check that
there are no non-vanishing F-terms of the type $(l_{7}\overline{l}_{X},l_{X}%
\overline{l}_{4},\phi _{X})(l_{7}\overline{l}_{4})^{m}.$ This is in
principle straightforward to do using the rules of eqs(\ref{rule1}) to
(\ref{rule4}).  In practice it is necessary to use algebraic computer programme to
do the search and the results quoted below have been obtained using MAPLE.
We find there is only one coupling of the dangerous type namely 
\begin{equation}
l_{7}{\overline{l}}_{4}\phi _{15},  \label{trilinear}
\end{equation}
which apparently spoils the flatness of the ($l_{7})_{N},$ ($\overline{l}%
_{4})_{N}$ direction. However there are three ways this conclusion may be
avoided.

\begin{itemize}
\item  The first possibility applies if there are couplings of higher order
of the type 
\begin{equation}
\frac{1}{M^{2(n-1)}}(l_{7}{\overline{l}}_{4})^{n}\phi _{15}
\end{equation}
One can easily check that the symmetries allow terms with $n=3p+1$ giving the
superpotential 
\begin{equation}
\phi _{15}(l_{7}{\overline{l}}_{4}+\frac{1}{M^{6}}(l_{7}{\overline{l}}%
_{4})^{4}+\frac{1}{M^{12}}(l_{7}{\overline{l}}_{4})^{7}+...)  \label{discr}
\end{equation}
We expect such terms to arise in the effective theory descending from the
string with $M$ given by the string scale. This means that although there is
no absolute flat direction for $l_{7}$ and $\overline{{l}}{_{4}}$, there
still exist a discrete zero of potential for non-zero value of $<l_{7}%
\overline{l}_{4}>=O(M^{2}).$ This is interesting because it corresponds to a
vacuum solution not continuously connected to the original Calabi Yau vacuum.

\item  The second way the ($l_{7})_{N},$ ($\overline{l}_{4})_{N}$ direction
may be flat is through a cancellation of its contribution to $F_{\phi _{15}}$
through other fields acquiring vevs.In particular one finds the following
terms 
\begin{equation}
\phi _{3}\phi _{4}\phi _{15},
\end{equation}
\begin{equation}
\phi _{1}\phi _{2}\phi _{15},
\end{equation}
\begin{equation}
\phi _{6}\phi _{5}\phi _{15},
\end{equation}
which we can use to give a non-zero VEV to $F_{\phi _{15}}$. This can happen
by adjusting VEVs of some of these singlet fields in such a way that the
whole term multiplied by $\phi _{15}$ in 
\begin{equation}
\phi _{15}(l_{7}{\overline{l}}_{4}+\phi _{3}\phi _{4}+\phi _{1}\phi
_{2}+\phi _{6}\phi _{5})
\end{equation}
vanishes. We cannot use $\phi _{3}$ and $\phi _{4}$ because it would
introduce a further non-zero F term through the allowed coupling $l_{7}{%
\overline{l}}_{3}\phi _{3}$. There is no such coupling for the fields $\phi
_{1}$ and $\phi _{2}$ so we consider whether there are new non-zero F terms
introduced in higher order if we allow them to acquire vevs. The dangerous
couplings are of the type 
\begin{equation}
(l_{7}{\overline{l}}_{4})^{n}\phi _{1}^{m}\phi _{2}^{k},
\end{equation}
\begin{equation}
(l_{7}{\overline{l}}_{4})^{n}\phi _{1}^{m}\phi _{2}^{k}\phi _{x},
\end{equation}
\begin{equation}
(l_{7}{\overline{l}}_{4})^{n}\phi _{1}^{m}\phi _{2}^{k}l_{7}{\overline{l}}%
_{x},
\end{equation}
\begin{equation}
(l_{7}{\overline{l}}_{4})^{n}\phi _{1}^{m}\phi _{2}^{k}l_{x}{\overline{l}}%
_{4}.
\end{equation}

Using MAPLE one can check that in fact the only allowed couplings belonging
to this set are $(l_{7}{\overline{l}}_{4})^{n}\phi _{1}^{m}\phi _{2}^{k}\phi
_{15},$for some specific $n$, $m$ and $k$. Including them the most general
superpotential involving \ $l_{7},$ $\overline{l}_{4},\phi _{1}$ and $\phi
_{2}$ are 
\begin{eqnarray}
\phi _{15}[l_{7}{\overline{l}}_{4} &+&(l_{7}{\overline{l}}_{4})^{4}+(l_{7}{%
\overline{l}}_{4})^{7}+  \nonumber \\
&+&{\rm (higher \; order \; terms \; involving \; only} \ l_{7} \;
{\rm and} \; {\overline{
l}}_{4})+  \nonumber \\
&+&\phi _{1}\phi _{2}  \nonumber \\
&+&{\rm (higher \; order \; terms \; involving \; only \;}\phi _{1}
\;{\rm and} \; \phi
_{2})+  \nonumber \\
&+&l_{7}{\overline{l}}_{4}\phi _{1}\phi _{2}+(l_{7}{\overline{l}}%
_{4})^{2}\phi _{1}\phi _{2}+(l_{7}{\overline{l}}_{4})^{2}\phi _{1}^{2}\phi
_{2}^{2}+  \nonumber \\
&+&{\rm  (higher \; order \; terms \; involving \; only} \; l_{7},
\; {\overline{l}}_{4}, \; \phi
_{1} \; {\rm and } \; \phi _{2})].
\end{eqnarray}

Clearly the vanishing of $F_{\phi _{15}}$ implies a relation between $<l_{7}%
\overline{l}_{4}>$ and $<\phi _{1}\phi _{2}>$, but leaves a flat direction
corresponding to the magnitude of $<l_{7}\overline{l}_{4}>.$

\bigskip

\item  Finally, the third option involves the moduli fields. The moduli, $m,$
determine the strength of the coupling in eq(\ref{trilinear}) so we have the
coupling 
\[
\phi _{15}(\lambda (m)l_{7}{\overline{l}}_{4}+\frac{\lambda ^{\prime }(m)}{%
M^{6}}(l_{7}{\overline{l}}_{4})^{4}+\frac{\lambda ^{\prime \prime }(m)}{%
M^{12}}(l_{7}{\overline{l}}_{4})^{7}+...) 
\]
Now $F_{\phi _{15}}$ can be made zero through the moduli fields acquiring
vevs to make the term in brackets vanish. Note that because this coupling
involves the field $\phi _{15}$ with vanishing vev this is the only
condition that needs be satisfied to ensure F-flatness. It is clear that if
the term {\it had} involved fields all of which acquire vevs this would not
have been true and all the moduli would have been determined by the
condition $F_{m_{i}}=0.$ We shall discuss this case later when we consider
directions which are only approximately flat.
\end{itemize}

This example has conveniently illustrated all the possibilities for
achieving flat directions. What we have shown is that there are equivalent
solutions to the compactified string theory in which the symmetry is broken
to $SO(10)$ through the $l_{7}$ and ${\overline{l}}_{4}$ vevs along the $N$
direction. Since these flat directions may have Planck scale vevs they are
of a different character to those hitherto investigated with vevs at
intermediate scales triggered by soft supersymmetry breaking mass terms.
These new solutions are on an equal footing to the original
compactifications with $E_{6}$ symmetry. Indeed what we have done is to
demonstrate that there is a richer moduli space describing the string vacuum
than just the string moduli discussed above. The disappointment the reader
may feel at yet another contribution to the degeneracy of string vacua
should be tempered by the fact that these string vacua may provide a more
promising starting point for a phenomenologically viable string theory. We
shall demonstrate this in the remainder of this paper investigating the
prospects for using these new solutions to build a realistic three
generation string theory.

\section{Low Energy string symmetries}

In this section we wish to explore the question whether we can use the
techniques discussed above to find a viable low energy three generation
theory with just the Standard Model gauge group at the compactification. The
starting point is the identification of the low-energy symmetries that must
be left unbroken until the electroweak breaking scale or the supersymmetry
breaking scale. In addition to the $SU(3)\times SU(2)\times U(1)$ gauge
symmetry there must be a symmetry capable of preventing rapid nucleon decay.
The latter \ may be the $Z_{2}$ R-parity of the MSSM or it may be a larger
discrete group. In addition there must be a further symmetry which prevents
one, and only one\footnote{%
This condition is necessary if the success of the gauge unification
predictions is to be maintained.}, pair of Higgs doublets from acquiring a
large invariant mass. In the MSSM this symmetry is absent and an arbitrarily
large $\mu $ term is allowed, so technically the MSSM is unnatural. In our
opinion such an unnatural tuning of a parameter is not acceptable in a
string theory because any term allowed by string symmetries typically occur
unsuppressed. Thus we require that the origin of the light Higgs be
guaranteed by the string symmetries. Moreover, as there are many more than a
single pair of Higgs doublets in the compactified string model, we have the
additional problem of ensuring that only one pair remains light.

\subsection{R-parity in the Calabi Yau 3 generation model.\label{sec:rparity}%
}

As we noted above the starting point for our string construction is the
three generation superconformal theory that is equivalent, in the large
radius limit, to a three generation Calabi-Yau theory. The gauge group of
this model can at most be broken to $SU(3)^{3}$ by Wilson lines. To break
this group further $E_{6}$ non-singlet fields must acquire vevs. It is
convenient to label the states of a $27$ by its $SU(3)^{3}$ transformation
properties. We have 
\begin{equation}
27=(1,3,\overline{3})\oplus (3,\overline{3},1)\oplus (\overline{3},1,3)
\end{equation}
where we take the first $SU(3)$ factor to be the colour group. Then the
leptons and Higgs states belong to the $(1,3,\overline{3})$ representation.
As discussed below the ``normal'' assignment is given by 
\begin{equation}
(1,3,\overline{3})=\left( 
\begin{array}{ccc}
H_{1} & H_{2} & l \\ 
E^{+} & \overline{\nu }_{R} & N^{0}
\end{array}
\right)  \label{su33rep}
\end{equation}
where $H_{1,2}$ are Higgs doublets, $l$ is a lepton doublet, $E^{+}$ is a
charged lepton, $\nu _{R}$ is a right-handed neutrino component and $N$ is
an $SO(10)$ singlet field (for the $SO(10)\otimes U(1)$ subgroup of $E_{6}).$
If we are to construct a deformation of the model with just the Standard
Model gauge group it is necessary for us to find flat directions which
involve fields transforming $\nu _{R}$ and as $N$ in eq(\ref{su33rep}).
Giving these vevs breaks the gauge group to $SU(3)\times SU(2)\times U(1)$
as required. Our task is to find flat directions which generate this
breaking but leave unbroken symmetries which prevent rapid nucleon decay and
which keep just one Higgs field light. It turns out that these requirements
largely determine the structure of the low-energy theory.

Consider the case that there is a $Z_{2}$ R-parity. The simplest possibility
is to identify it with a discrete symmetry of the string compactification.
However this turns out not to be consistent with the requirement of a single
Higgs pair of fields and three quark and lepton generations. To see this we
note that Higgs doublets are R-parity even and must come from R-parity even $%
(1,3,\overline{3})$ and $(1,\overline{3},3)$ multiplets in the underlying $%
SU(3)^{3}$ theory. Our assumption is that the R-parity commutes with the
gauge symmetry and thus all three doublets of an R-parity even multiplet of
eq(\ref{su33rep}) must be Higgs doublets ({\it not} as shown in
eq(\ref{su33rep})). Thus we see that a $(1,3,\overline{3})$ representation contains
two Higgs doublets with weak hypercharge +1 and one Higgs doublet with weak
hypercharge -1. Upon $(1,3,\overline{3})$ and $(1,\overline{3},3)$ breaking
some of the Higgs fields may acquire mass through the coupling of positive
and negative hypercharge states. It is clear that to have the possibility
that only one pair of Higgs fields with opposite hypercharge are left light
we must have 
\begin{equation}
N_{R+}-N_{\overline{R}+}=0  \label{nh1}
\end{equation}
where $N_{R+,\overline{R}+}$ are the number of positive R-parity states in
the $(1,3,\overline{3})$ and $(1,\overline{3},3)$ representations
respectively. Quark and leptons belong to R-parity odd states and in order
for there to be just three generations we must have 
\begin{equation}
N_{R-}-N_{\overline{R}-}=3  \label{nl1}
\end{equation}
Note that the ``three generation'' condition of the underlying Calabi-Yau
manifold, $N_{R+}+N_{R-}-N_{\overline{R}+}-N_{\overline{R}-}=3,$ does {\it %
not} guarantee three quark and lepton generations; this requires a
definition of quark and lepton which R-parity supplies. The only $Z_{2}$
symmetry in the model of Section \ref{sec:model} is the anticyclic
permutation of the three k=16 factors of the underlying superconformal
model. In this case only the fields $l_{4}-l_{5},$ $l_{8}-l_{9},$ and $%
\overline{l}_{5}-\overline{l}_{6}$ are odd under the $Z_{2},$ giving $%
N_{R+}=7,$ $N_{R-}=2,$ $N_{\overline{R}+}=5,$ and $N_{\overline{R}-}=1.$
Clearly this does {\it not} satisfy eqs(\ref{nh1}) and (\ref{nl1}) so we
cannot identify this $Z_{2}$ symmetry with R-parity.

\begin{table}[tbp] \centering%
%
\begin{tabular}{|l|l|}
\hline
$l_{4}l_{5}l_{3}$ & ${\overline{l}}_{4} {\overline{l}}_{4} {\overline{l}}_{2}$ \\
$l_{7}l_{7}l_{3}$ & ${\overline{l}}_{1} {\overline{l}}_{3} {\overline{l}}_{4}$ \\ 
$l_{1}l_{5}l_{6}$ & ${\overline{l}}_{5} {\overline{l}}_{6} {\overline{l}}_{2}$ \\ 
$l_{1}l_{4}l_{6}$  &  \\ 
$l_{2}l_{4}l_{6}$  &  \\ 
$l_{2}l_{5}l_{6}$  &  \\ 
$l_{2}l_{6}l_{7}$  &  \\ 
$l_{8}l_{9}l_{2}$  &  \\ \hline
\end{tabular}
\caption{ Tri-linear couplings for the matter fields.\label{Table:trilinear}}%
\end{table}%
%

\begin{table}[tbp] \centering%
%
\begin{tabular}{|l|l|}
\hline
$q_{3} q_{3} q_{1}$ & $l_{5} Q_{1} q_{2}$ \\ 
$q_{1} q_{2} q_{3}$ & $l_{6} Q_{3} q_{1}$ \\ 
$Q_{3} Q_{3} Q_{1}$ & $l_{6} Q_{1} q_{3}$ \\ 
$Q_{1} Q_{2} Q_{3}$ & $l_{5} Q_{3} q_{1}$ \\ 
$l_{3} Q_{3} q_{3}$ & $l_{4} Q_{1} q_{3}$ \\ 
$l_{2} Q_{2} q_{2}$ & $l_{1} Q_{3} q_{2}$ \\ 
$l_{2} Q_{3} q_{3}$ & $l_{1} Q_{2} q_{3}$ \\ 
$l_{4} Q_{2} q_{1}$ &  \\ \hline
\end{tabular}
\caption{ Tri-linear couplings involving the quark fields for the case the
phase group is modded out .\label{Table:trilinearq}}%
\end{table}%
%

\begin{table}[tbp] \centering%
%
\begin{tabular}{|l|}
\hline
{\ $l_{1}{\overline{l}}_{1}\phi _{3}$} \\ \hline
{\ $l_{1}{\overline{l}}_{5}\phi _{11}$} \\ \hline
{\ $l_{1}{\overline{l}}_{6}\phi _{10}$} \\ \hline
{\ $l_{2}{\overline{l}}_{1}\phi _{3}$} \\ \hline
{\ $l_{2}{\overline{l}}_{4}\phi _{7}$} \\ \hline
{\ $l_{2}{\overline{l}}_{5}\phi _{8}$} \\ \hline
{\ $l_{2}{\overline{l}}_{6}\phi _{9}$} \\ \hline
{\ $l_{3}{\overline{l}}_{1}\phi _{4}$} \\ \hline
{\ \ $l_{4}{\overline{l}}_{3}\phi _{3}$} \\ \hline
{\ $l_{4}{\overline{l}}_{5}\phi _{17}$} \\ \hline
{\ $l_{5}{\overline{l}}_{3}\phi _{3}$} \\ \hline
{\ $l_{5}{\overline{l}}_{6}\phi _{16}$} \\ \hline
{\ $l_{6}{\overline{l}}_{1}\phi _{20}$} \\ \hline
\end{tabular}
\begin{tabular}{|l|}
\hline
{\ $l_{6}{\overline{l}}_{2}\phi _{12}$} \\ \hline
{\ $l_{6}{\overline{l}}_{3}\phi _{4}$} \\ \hline
{\ $l_{7}{\overline{l}}_{3}\phi _{3}$} \\ \hline
{\ $l_{7}{\overline{l}}_{4}\phi _{15}$} \\ \hline
{\ $l_{8}{\overline{l}}_{1}\phi _{18}$} \\ \hline
{\ $l_{8}{\overline{l}}_{2}\phi _{13}$} \\ \hline
{\ $l_{8}{\overline{l}}_{3}\phi _{1}$} \\ \hline
{\ $l_{8}{\overline{l}}_{3}\phi _{16}$} \\ \hline
{\ $l_{9}{\overline{l}}_{1}\phi _{19}$} \\ \hline
{\ $l_{9}{\overline{l}}_{2}\phi _{13}$} \\ \hline
{\ $l_{9}{\overline{l}}_{3}\phi _{2}$} \\ \hline
{\ $l_{9}{\overline{l}}_{3}\phi _{5}$} \\ \hline
\\ \hline
\end{tabular}
\caption{Terms of the type $l \bar{l} \phi$.\label{Table:trilinear1}}%
\end{table}%
%

Since there are several $Z_{3}$ symmetries it may be possible to use one to
generate a baryon parity \cite{ibanezross}. We will investigate this
possibility elsewhere but here we wish to point out a novel possibility
which utilises an approximate $Z_{2}$ symmetry to implement an approximate
R-parity. For example in Table \ref{Table:trilinear} one sees that $l_{8}$
and $l_{9}$ couple only in pairs meaning that these couplings are invariant
under an effective $Z_{2}$ symmetry under which $l_{8,9}$ are odd. The point
is that the $Z_{3}$ symmetries give rise to an effective $Z_{2}$ symmetry in
the trilinear couplings which may be used to define the R-parity.

With this motivation we consider whether any of these approximate symmetries
are capable of generating such an R symmetry. Since the discrete gauge group
factor remains the same the constraints of eqs(\ref{nh1}) and (\ref{nl1})
still apply. It is straightforward to enumerate all possible $Z_{2}$
symmetries consistent with the terms appearing in \ref{Table:trilinear}. For
example $l_{3}$ is necessarily even while $l_{8,9}$ may both be odd or both
be even etc. However in all cases the number of $Z_{2}$ odd states is even
in both the $(1,3,\overline{3})$ and $(1,\overline{3},3)$ sectors. Thus eq(%
\ref{nl1}) cannot be satisfied and even the approximate symmetries cannot be
identified with R-parity.

Luckily there is another way to implement this approxmate R-parity. The
problem we have encountered follows because the R-symmetry has been assumed
to commute with $SU(3)^{3}$ leading immediately to eqs(\ref{nh1}) and (\ref
{nl1}). If, however, we identify R-parity with CU where C is a $Z_{2}$
discrete symmetry and U is the $Z_{2}$ discrete gauge group factor given by
the $SU(2)_{L}\otimes SU(2)_{R}$ diagonal group elements$(-1,-1,1)_{L}%
\otimes (-1,-1,1)_{R}$ the counting changes. In this case, cf eq(\ref
{su33rep}), a $C+$ $(1,3,\overline{3})$ state contains two R-parity even
doublets which are to identified with Higgs doublets carrying opposite weak
hypercharge together with a R-parity odd doublet which is to be identified
with a lepton doublet. The identifications are reversed for a $C-$ state.
With similar identifications for the $(1,\overline{3},3)$ states we find the
conditions of eqs(\ref{nh1}) and (\ref{nl1}) are replaced by

\begin{eqnarray}
N_{C+}-N_{\overline{C}+} &=&3  \label{nlh2} \\
N_{C-}-N_{\overline{C}-} &=&0  \nonumber
\end{eqnarray}

Consider first the possibility we identify $C$ with the $Z_{2}$ given by the
anticyclic permutation of the three k=1 factors of the underlying
superconformal model. This corresponds to $N_{C+}=7,$ $N_{\overline{C}+}=5,$ 
$N_{C-}=2,$ $N_{\overline{C}-}=1.$ In this case the constraints of eqs(\ref
{nh1}) and (\ref{nl1}) are not satisfied. Of course one may go to higher
dimension discrete symmetries and this indeed is a possibility given that
there are several $Z_{3}$ symmetries of the theory. We shall investigate
this possibility in another paper but here we wish to explore the novel
possibility discussed above which utilises an approximate $Z_{2}$ symmetry
to implement an approximate R-parity. In this case it is possible to satisfy
these equations through the identification of $C$ with the approximate $%
Z_{2} $ symmetry under which $l_{8,9}$ and $\overline{l}_{1,3}$are odd while
all other states are even. For it $N_{C+}=7,$ $N_{\overline{C}+}=4,$ $%
N_{C-}=2,$ $N_{\overline{C}-}=2,$ satisfying eqs(\ref{nlh2}). The symmetry
clearly leaves the terms of Table \ref{Table:trilinear} invariant. Provided
the flat direction has non-zero vevs for R-parity even states only R-parity
will be preserved. This happens if \hbox{$<$}$N>,$ \hbox{$<$}$\overline{N}>$
belong to $C$ even and \hbox{$<$}$\nu _{R}>,$ \hbox{$<$}$\overline{\nu }%
_{R}> $ belong to $C$ odd states only. However this, by itself, is not
sufficient to guarantee that R-symmetry is preserved because higher
dimension terms may become important along these flat directions and these
may not respect the approximate $Z_{2}$ symmetries. Thus it is important to
identify the origin of this approximate symmetry so that one may be sure
that it is not broken when fields develop vevs along flat directions. From
Tables \ref{Table:1} and \ref{Table:2} we see that the states
$l_{8,9}$ and $\overline{l}_{1,3}$ are distinguished by being the only
states with charge $q_{i},$ $%
\overline{q}_{i}=2$ $\overline{q}_{i}$ $Mod$ $6$ ,$i=2,3,4.$ The approximate 
$Z_{2}$ is due to the underlying $Z_{3}$ symmetries associated with these
charges. The flat direction we explore in detail here preserves has all $%
SU(3)^{3}$ singlet combinations carrying zero $\overline{q}_{i}$ charge $\
Mod$ $6$, so it is this $Z_{3}$ (called $Z_{3}^{G}$ henceforth$)$ that
maintains the approximate $Z_{2}.$ The trilinear couplings $l^{3}$ allowed
by this symmetry have $\overline{q}_{i}$ charge $0.0.4$, $0.2.2,$ and $4.4.2$%
. This implies a potential conflict with the need to obtain a $Z_{2}$
symmetry because the assignment assumed above assigned states carrying
charge $0$ and $4$ positive $Z_{2}$ charge while states carrying charge $2$
had negative $Z_{2}$ charge. This assignment is inconsistent with a coupling 
$4.4.2.$ In Table \ref{Table:trilinear} this troublesome coupling is absent
but one must check whether it is generated at an unacceptable level when
large vevs develop along flat directions. For a coupling to be allowed

\begin{equation}
\sum_{j}\overline{q}_{i}^{j}=16,i=2,3,4  \label{charge}
\end{equation}
where $\overline{q}_{i}^{j}$ is the $\overline{q}_{i}$ charge for the jth
field. From Table \ref{Table:1} we see that the trilinear term $%
l_{3}l_{6}l_{8}$ can satisfy this condition. It does not appear in Table \ref
{Table:trilinear} because the other charge constraints corresponding to
other symmetries are not satisfied. However if all other discrete symmetries
are broken at some stage this term will be allowed. Does it lead to rapid
nucleon decay? In fact it does not because the superfield $l_{3}$ is a
permutation singlet and hence does not contain quark components. As a result
the R-parity violating coupling is only allowed in the lepton sector
generating lepton number violation but not baryon number violation.

The only $4.4.2$ coupling which {\it does} involve quarks is $%
l_{6}l_{6}l_{8.9}.$ However this coupling is forbidden by the constraint of
eq(\ref{charge}). The simplest way to see this is to note that $\sum_{i,j}%
\overline{q}_{i}^{j}=54$ instead of $48$ required by eq(\ref{charge}). We
must also check that this coupling does not appear together with the $%
SU(3)^{3}$ singlet combinations of fields acquiring large vevs so that it
does not appear when the theory is deformed along a flat direction. This
will be the case for the phenomenologically interesting direction discussed
below, but due to a combination of the constraints of eq(\ref{charge}).

In conclusion we have investigated the possibility of preventing rapid
nucleon decay via discrete symmetries in the three generation Gepner model.
We have found that it is possible to ensure stability to the required order
through a new, approximate, $Z_{2}$ symmetry involving the quark sector.
This symmetry allows dimension 4 lepton number violating terms and is thus
different from the normal R-parity of the MSSM. In fact it is a version of
``Baryon'' parity \cite{ibanezross}. Of course suppression of the dimension 4
terms leading to nucleon decay may not be sufficient because dimension 5
terms can still occur at an unacceptable rate. These terms require a
detailed model and we will discuss them once we have filled in some of these
details.

\subsection{Light Higgs doublets\label{sec:higgs}}

While our discussion of R-parity allowed for the possibility that there
should be left light just one pair of Higgs doublets with opposite weak
hypercharge, it did not guarantee that this should be the case. However it
is important that a symmetry should ensure the lightness of the Higgs states
because the Higgs doublets can acquire an $SU(3)\otimes SU(2)\otimes U(1)$
invariant mass and the symmetry is needed to prevent this happening, without
fine tuning of parameters, when the underlying $SU(3)^{3}$ symmetry is
broken. To address this question we will try to identify an effective $Z_{2}$
symmetry involving the chiral Higgs supermultiplets which will generate a
chirality ensuring that one Higgs pair can not acquire a mass. We denote by $%
N_{H_{1,2}^{\pm }},$ and $N_{\overline{H}_{1,2}^{\pm }}$ the number of Higgs
fields with $Z_{2}$ ``chirality'' $\pm 1$ coming from the $(1,3,\overline{3}%
) $ sector and the $(1,\overline{3},3)$ sector respectively. The definition
of ``chirality'' here requires that masses involve a coupling of a positive
to a negative chirality state.The requirement that there should be just one
pair of Higgs doublets forbidden by this chirality from acquiring a mass is
then given by eq(\ref{nlh2}) together with 
\begin{equation}
N_{H_{1}^{+}}-N_{H_{1}^{-}}+N_{\overline{H}_{2}^{+}}-N_{\overline{H}%
_{2}^{-}}=1  \label{1higgs}
\end{equation}
It is now straightforward to check whether there is a symmetry in the
compactified string model capable of satisfying this constraint. Consider
the allowed mass terms consistent with the $Z_{3}^{G}$ symmetry which
ensures nucleon stability. A Higgs mass term can be generated by $l^{3}$ or $%
\overline{l}^{3}$couplings or by a $l\overline{l}$ coupling. As discussed
above The $l^{3}$ couplings allowed by the $Z_{3}^{G}$ symmetry have $%
\overline{q}_{i}$ charge $0.0.4$, $0.2.2,$ and $4.4.2.$ The discussion is
made more complicated because, as mentioned above, the $4.4.2$ term violates
lepton number and so mixes Higgs and leptons. Let us first ignore this term,
in which case there is a well defined separation of Higgs and leptons as
discussed above eq(\ref{nlh2}). Consider first the mass terms generated by
the \hbox{$<$}$N>$ vev.Such a $l^{3}$ term will generate a mass term for an $%
H_{1,2}$ pair coupling $\overline{q}_{i}$ charge $0$ to $4$ states and $%
\overline{q}_{i}$ charge $2$ to $2$ states. Thus a consistent assignment of
the $Z_{2}^{G}$ ``chirality'' is to make all the $0$ states positive and all
the $4$ states negative. The $0.2.2$ mass terms generated by a $<\nu _{R}>$
vev in the $2$ direction is consistent with this $Z_{2}^{G}$ ``chirality''
if the Higgs state in the $2$ multiplet is odd. The mass term arising from
the $0.2.2$ coupling with a \hbox{$<$}$N>$ vev in the $0$ multiplet involves
lepton states only and plays a role in the determination of the lepton
parities as discussed below. We turn now to the mass terms generated by $%
\phi l\overline{l}$ couplings. Note that the mass terms generated by a $\phi 
$ vev involve $0.0$, $2.\left( -2\right) $ and $4.(-4)$ $\overline{q}_{i}$
charges. Thus in the $\overline{l}$ sector we must assign all $0$ Higgs
states {\it negative} and all the -$4$ and $-2$ states {\it positive} $%
Z_{2}^{G}$ parity. Finally consideration of the $\overline{l}^{3}$ couplings
shows that they are consistent with these assignments. The resulting
assignments for the ``Higgs'' states are shown in Table \ref{Table:parity}.

\begin{table}[tbp] \centering%
%
\begin{tabular}{|l|l|}
\hline
$O$ : $\left( 
\begin{array}{ccc}
\overline{H}_{+} & H_{+} & l_{+} \\ 
E_{+} & \overline{\upsilon }_{R+} & N_{+}
\end{array}
\right) $ & $\overline{O}$ : $\left( 
\begin{array}{ccc}
H_{-} & \overline{H}_{-} & \overline{l}_{-} \\ 
\overline{E}_{-} & \upsilon _{R-} & \overline{N}_{-}
\end{array}
\right) $ \\ 
$4$ :$\left( 
\begin{array}{ccc}
\overline{H}_{-} & H_{-} & l_{+} \\ 
E_{+} & \overline{\upsilon }_{R+} & N_{+}
\end{array}
\right) $ & $\overline{4}$ : $\left( 
\begin{array}{ccc}
H_{+} & \overline{H}_{+} & \overline{l}_{-} \\ 
\overline{E}_{-} & \upsilon _{R-} & \overline{N}_{-}
\end{array}
\right) $ \\ 
$2$ : $\left( 
\begin{array}{ccc}
\overline{l}_{-} & l_{+} & H_{-} \\ 
H_{-}^{+} & H_{+}^{0} & E_{+}^{0}
\end{array}
\right) $ & $\overline{2}$ : $\left( 
\begin{array}{ccc}
l_{+} & \overline{l}_{-} & \overline{H}_{+} \\ 
\overline{H}_{+}^{-} & \overline{H}_{-}^{0} & \overline{E}_{-}^{0}
\end{array}
\right) $ \\ \hline
\end{tabular}
\caption{
$Z_2^G$ chiralities of the Higgs and lepton states. Here $\bar{H},H$ have the quantum numbers of the 
Higgs doublets $H_{1,2}$ with opposite weak hypercharge. The Higgs states $\bar{H},H$ can mix with the lepton states 
$\bar{l},l$.\label{Table:parity}}%
\end{table}%
%

For the leptons the assignment of parity follows using the same general
argument and the results are also shown in Table \ref{Table:parity}. At this
point we may include the coupling between the ``Higgs'' states and the
``lepton'' states arising from the $2.4.4$ with a $<\nu _{R}>$ vev in the $2 
$ direction. The associated mass terms are consistent with the assignments
of \ref{Table:parity} (in fact the relative sign between the leptons and
quarks is determined by this coupling to be that given in \ref{Table:parity}%
). It is now straightforward to determine the number of states which must be
left light due to chirality. We have $N_{H_{1+}}=N_{0}+N_{\overline{4}}+N_{%
\overline{2}}=8,$ $N_{H_{2-}}=N_{\overline{0}}+N_{4}+N_{2}=7$ and so there
are $N_{H_{1+}}-N_{H_{2-}}=1$ Higgs doublets $H_{1}$ left light. Similarly
we find $N_{H_{2+}}-N_{H_{1-}}=1$ Higgs doublets $H_{2}$ left light.
Similarly we find $N_{L+}=N_{0}+N_{4}+N_{2}+N_{\overline{2}}=11,$ $N_{%
\overline{L}-}=N_{\overline{0}}+N_{\overline{4}}+N_{\overline{2}}+N_{2}=8$
so the number of lepton doublets left light is $N_{L+}-N_{\overline{L}-}=3$
as desired. One may check that the $4.4.2$ couplings which mix Higgs and
leptons do not disturb this counting for using the chiralities of Table \ref
{Table:parity} which were determined including this coupling we find $%
N_{H_{2+}}+N_{L+}-N_{H_{1-}}-N_{\overline{L}-}=4$ and $%
N_{H_{1+}}-N_{H_{2-}}=1$ corresponding to three ``lepton'' and one ``Higgs''
doublet which mix and one Higgs doublet of the opposite hypercharge.
Remarkably this demonstrates that the {\it same} symmetry that is
responsible for stabilising the nucleon also ensures three generations of
leptons and one light pair of Higgs states remains light.

\section{A Three generation $SU(3)\otimes SU(2)\otimes U(1)$ Compactified
String Theory\label{sec:3genCY}}

Having identified the string symmetries we need to remain unbroken to the
supersymmetry breaking scale we are now in a position to determine whether
we can obtain a compactified string theory with these symmetries by
deforming the three generation Gepner model along absolutely flat
directions. The breaking along the $N$ direction should be $C$ even to
implement the approximate R-parity discussed in Section \ref{sec:rparity}%
.They should also be in the even $Z_{2}^{G}$ parity sector to maintain a
light Higgs state as discussed in Section \ref{sec:higgs}. Further, as
discussed in Section \ref{sec:flat}, absolutely flat directions require a
discrete R-symmetry. Consider the discrete symmetry associated with the
charge $\alpha _{1}$ as defined in eq(\ref{alpha})$.$ Terms allowed by this
symmetry in the superpotential must have charge +1 and so there are no terms
involving charge 0 fields on their own. As a result (cf. Section \ref
{sec:flat}), these are likely to be absolutely flat directions. We thus
consider the $C$ even, $Z_{2}^{G}$ parity even, $\alpha _{1}$ neutral states 
$l_{4,5,7}$ and $\overline{l}_{4,5,6}$ as candidate states to acquire
non-vanishing vevs along the $N$ direction. We concentrate here on the
possibility only $l_{7}$ and $\overline{l}_{4}$ acquire vevs in the $N$
direction and postpone a discussion of the other possibilities to a future
publication. Turning to the breaking along the $\nu _{R}$ direction we
follow the logic just discussed and conclude this should be along a $C$ odd, 
$Z_{2}^{G}$ parity odd, $\alpha _{1}$ neutral direction. In this case there
is no choice and it must in the $l_{8},$ $\overline{l}_{1}$ direction.
However this direction is not D-flat due to the charge, $\overline{\alpha }%
_{1},$ carried by these fields. To obtain a flat direction we allow $l_{2}$
to acquire a vev along the $N$ direction. Since it has the opposite sign of
charge under the $\overline{U}_{1}$ its vev will adjust to cancel this
D-term.

It is straightforward to check, following the same discussion as in Section 
\ref{sec:flatex}, that the combination of fields $l_{8}|_{\nu _{R}},$ $%
\overline{l}_{1}|_{\nu _{R}},$ $l_{7}|_{N}$, $l_{2}|_{N\hbox{ }}$and $%
\overline{l}_{4}|_{N}$ is indeed flat to all orders. The only gauge
invariant combination of fields has the form $(l_{2}l_{\overline{4}}l_{8}l_{%
\overline{1}})^{n}(l_{7}l_{\overline{4}})^{m}.$ This has the charge
structure $(1,16,4,10)^{n}(0,12,12,12)^{m}$ where we have denoted the
charges $(3\alpha _{1},18\alpha _{i},i=2,3,4)$ and we must allow symmetric
permutations in the last three charges. Clearly eq(\ref{rule1}) requires $%
n=3p+1$ for some integer $p.$ Inserting this and applying eq(\ref{rule1})
again gives 
\begin{eqnarray}
16a+4b+10c+12m &=&18r+16 \\
16b+4c+10a+12m &=&18s+16  \nonumber \\
16c+4a+10b+12m &=&18t+16  \nonumber \\
a+b+c &=&3p+1  \nonumber
\end{eqnarray}
where $a,b,c,m,r,s,t,p\in Z.$ The appearance of the separate terms involving 
$a,b,c$ follows from the possibility of making cyclic permutations of
charges. It is straightforward to check that these equations have no
solution to the term $(l_{2}l_{\overline{4}}l_{8}l_{\overline{1}%
})^{n}(l_{7}l_{\overline{4}})^{m}$ is not allowed. In this case the
R-symmetry guaranteeing its absence arises as a combination of the discrete
symmetries and not just the $\alpha _{1}$ charge as before. This provides an
example of a flat direction breaking $E_{6}$ to $SU(5)$ just as is desired.
Allowing for Wilson line breaking the gauge group can be further broken to
just the $SU(3)\otimes SU(2)\otimes U(1)$ of the S$\tan $dard Model.

\subsection{The spectrum at the compactification scale}

It is straightforward now to determine the spectrum in the model allowing
for the string scale vevs along the flat direction just discussed and the
Wilson line breaking that follows if they are associated with the freely
acting group of cyclic permutations that is modded out when forming the 3
generation model. After Wilson line breaking the leptons belong to the
permutation symmetric states, the left-handed quark states belong to the
states transforming as $e^{i2\pi /3}$, and the right-handed quark states
belong to the states transforming as $e^{i4\pi /3}$ under cyclic
permutations. In order to avoid a confusing profusion of indices we will not
make explicit the permutation symmetry in the discussion below.

The Higgs doublets left light are the $H_{1,2}$ pairs of doublets in the $C$%
-even states $l_{1,4,5}$ the $H_{3}$ doublets in the $C$-odd states $l_{8,9}$
and the $\overline{H}_{3}$ doublets in the $C$-odd states $\overline{l}%
_{1,3}.$ The remaining Higgs doublets acquire mass by their coupling to the $%
N,\overline{N}$ vevs via ${\bf 27}^{3}$ or ${\bf \overline{27}}^{3}$
or ${\bf 27.\overline{27}%
.1}$ couplings.

The lepton doublets left light are in the $C-$even states $l_{1,2,3,4,5,6}$
and $\overline{l}_{2,5,6}.$

Finally there are $C$-even $SU(2)_{L}$ and $SU(2)_{R}$ quark doublets left
light in the multiplets $l_{1,2,4,5,6}$ and $\overline{l}_{5,6}.$ In
addition there are left light $C-$odd $SU(2)_{L}$ and $SU(2)_{R}$ quark
doublets in $l_{9}$ and $\overline{l}_{3}.$ There are also $SU(2)_{L}$ and $%
SU(2)_{R}$ quark singlets left light in the $C$-even multiplets $l_{1,4,5}.$

It is possible to give vevs to further fields without disturbing the flat
direction. For example the field $\phi _{16}$ may acquire a vev giving a
mass to the quark and lepton doublets in the multiplets $l_{5},$ $\overline{l%
}_{6}.$ It is clear that a complete discussion of the various possibilities
require a careful analysis of the many possible flat directions and this is
beyond the scope of the present paper. Here we wish to demonstrate that the
methods developed above offer a rich building kit for generating
phenomenologically interesting models. To complete this programme we take
the model thus developed, including the $\phi _{16}$ vev, and show that
symmetry breaking below the compactification scale can lead to a viable
model.

\subsection{Intermediate scale breaking}

As we have seen the breaking at the compactification scale leaves a model
with the required $SU(3)\otimes SU(2)\otimes U(1)$ gauge group but with many
additional states beyond the MSSM spectrum lying in vectorlike
representations with respect to $SU(3)\otimes SU(2)\otimes U(1).$ It turns
out it is not possible to give all these additional states a mass through
vevs developing along absolutely flat directions. However it is known that
radiative corrections can (and in fact often do) drive scalar masses
negative triggering further breaking. In general this will not occur along
absolutely flat directions so the breaking typically occurs at scales
beneath the compactification scale but far above the electroweak scale. Such
breaking can readily give mass to the additional states as we now
demonstrate. Let us consider whether the flat direction discussed above can
be extended to include the scalar fields $\phi _{8},$ $\phi _{11},\phi
_{16}. $ In this case one may find a superpotential term allowed by all the
symmetries of the form $\lambda (m)(l_{2}l_{\overline{4}}\phi _{8}\phi
_{16})^{2}l_{7}l_{\overline{4}}/M^{7}$. As discussed in Section \ref
{sec:flat} the appearance of this term means the theory is not F-flat.
However large vevs may still develop along the $\phi _{8},$ $\phi _{16}$%
directions if the sum of the soft supersymmetry breaking masses squared for
these fields, $m_{\phi _{16}}^{2}+m_{\phi _{8}}^{2}$ is negative. Moreover 
 \cite{rossdumitru} radiative corrections involving the Yukawa couplings of
these fields is very likely to drive these mass squared negative so breaking
along these field directions is quite reasonable. Assuming this does indeed
happen it is straightforward to determine the scale of this breaking. Given
string scale vevs for the fields $l_{8}|_{\nu _{R}},$ $\overline{l}%
_{1}|_{\nu _{R}},$ $l_{7}|_{N}$, $l_{2}|_{N\hbox{ \ }}$and $\overline{l}%
_{4}|_{N}$ develop along the absolutely flat direction as discussed in
Section \ref{sec:3genCY} the largest F-term are the ones which reduce the
power of $\phi _{8},$ $\phi _{16}$ since these acquire vevs below the Planck
scale, i.e. $F_{\phi _{8},\phi _{16}}.$ However these terms may be made much
smaller by adjusting the moduli field vevs, $<m>($it is at the stage of
intermediate scale breaking that we expect the string moduli vevs to be
fixed). As a result the F-terms may all be made of the same order as $F_{m}$
or $F_{l_{2},l_{7,}l_{\overline{4}}}.$ These are of order ($\phi _{8}\phi
_{16})^{2}/M^{2}$ leading to the potential 
\begin{equation}
V=m_{\phi _{16}}^{2}\phi _{16}{}^{2}+m_{\phi _{8}}^{2}\phi _{8}^{2}+O((\phi
_{8}\phi _{16})^{4}/M^{4})
\end{equation}
Minimisation of this potential for $m_{\phi _{16}}^{2}+m_{\phi _{8}}^{2}$
negative leads to the vevs $\phi _{16}\approx \phi _{8}=O(((m_{\phi
_{16}}^{2}+m_{\phi _{8}}^{2})M^{4})^{1/6})\approx 10^{13}GeV$ where we have
used a supersymmetry breaking scale of $O(1TeV).$ In a similar manner we may
deduce that $\phi _{11}$ may acquire a vev of the same order. Allowing for
such breaking below the compactification scale we find that all the masses
allowed by the residual symmetries of the model are generated. In particular
all vectorlike states with respect to these residual symmetries acquire
intermediate scale masses. From Table \ref{Table:singlets} we see that the
fields $\phi _{8},$ $\phi _{11},\phi _{16}$ all have zero $\overline{\alpha }%
_{2,3,4}$ charge $Mod$ $6.$ This means they may acquire vevs without
breaking the $Z_{3}^{G}$ symmetry of Section \ref{sec:higgs} and \ref
{sec:rparity}. As a result we are guaranteed to have three generations of
leptons and one pair of Higgs doublets left light after intermediate scale
breaking.

\subsection{Nucleon decay}

After intermediate scale breaking we arrive at a low energy supersymmetric
theory with just the Standard Model gauge group and with the MSSM\ minimal
matter content. However the theory is not identical to that of the MSSM
because, as discussed above, there is a Baryon parity stabilising the
nucleon which allows lepton-number-violating dimension 4 terms in the
Lagrangian. At this stage we are in a position to discuss the dimension 5
contributions to nucleon decay. There are two dimension 5 operators, namely
$%
QQQl|_{F}$ and $Q_{1}^{c}Q_{1}^{c}Q_{1}^{c}E|_{F}.$ The potential problem
with dimension 5 operators follows because, although suppressed by the
inverse of a large mass, the suppression of them is typically inadequate by
itself to increase the nucleon lifetime to be consistent with experimental
bounds. However the second operator turns out to have a very strong
suppression due to mixing angles and the suppression factor that occurs when
dressing the operator to change the sfermions to fermions. Thus we
concentrate here on the former operator. The first point to note is that
these operators are absent before vevs develop along flat directions due to
the underlying gauge symmetry because there is no{\bf \ }${\bf
27.27.27.27}$%
{\bf \ }coupling. However once we allow large vevs for fields in the $%
{\bf 27}$ representations these terms can arise through tree level graphs
giving the higher dimensional couplings $\frac{1}{M^{2}}{\bf
27.27.27.27.<27}%
^{\dagger }{\bf >|}_{F}.$ The magnitude of these terms depends on the
effective mass scale, $M_{eff}^{-1}=\frac{{\bf <27}^{\dagger }{\bf
>}}{M^{2}}
$, associated with the operator. Even if one uses a Planck scale mass
suppression, $M_{eff}=M_{Planck}$, after allowing for gaugino dressing of the
operator and estimating the operator matrix element of the resulting four
fermion operator, one needs a further suppression of a further factor of $%
10^{-5}$ in amplitude. Thus there must be some approximate symmetry left by
the vevs developing along flat directions which will guarantee the smallness
of $(M_{eff}/M_{Planck})^{-1}$. Hovever, in contrast to the case of the
symmetry protecting against dimension 4 operators, this approximate symmetry
may be broken far above the electroweak scale because the necessary
suppression factor is much smaller. It is straightforward to check that our
Baryon parity does not by itself forbid such terms. For example for the case
in which we mod out by a phase twist the couplings of Table \ref
{Table:trilinear} generate the term $Q_{1}Q_{2}Q_{3}l_{2}<l_{2N}^{\dagger
}>$
through the exchange of the third component of $Q_{2}$ which acquires a mass
through $<l_{2N}>$ . Thus in this case $M^{2}\approx |<l_{2N}^{\dagger }>$ %
\hbox{$\vert$}$^{2}$ so the dimension 5 operator $Q_{1}Q_{2}Q_{3}l_{2}$ is
suppressed by $1/<l_{2N}>.$ Allowing for the (s)quark mixing angles going
from current to light quark mass eigenstates gives a factor of approximately
($\sin \theta _{c})^{2}$ (assuming mixing angles in the squark sector are
comparable to the quark sector) leaving a further suppression of $10^{-4}$
still needed. Here we consider two effects that may be responsible for the
remaining suppression.

\subsubsection{Quark and lepton mixing angle suppression}

One possible source of suppression is that there is a small mixing angle
associated with the lepton doublet $l_{2}.$ As we have discussed the lepton
sector is very rich so the determination of the mixing angles is quite
involved. In particular, as discussed in more detail in the next Section,
the lepton doublets may be mixtures of the lepton doublets in the fields $%
l_{1,2,3,4,5,6,7,8,9}$ and $\overline{l}_{1,3}$ and also of the ``Higgs''
doublet in $\overline{l}_{2}.$ In the case of the operator $%
Q_{1}Q_{2}Q_{3}l_{2}$ there is a reason why we expect the mixing angle to be
very small, much less than the minimum required. This follows because of the
coupling (cf Table \ref{Table:trilinear}) $l_{2}l_{8}l_{9}$. The lepton
doublet in $l_{2}$ acquires a large mass from the term $l_{2}l_{9}<l_{8}|_{%
\nu _{R}}>$. As a result the component of $l_{2}$ in the light lepton
doublets may be extremely small, the mixing angle being suppressed by the
factor $1/<l_{8}|_{\nu _{R}}>$. This can readily suppress the contribution
to nucleon decay of the dimension 5 operator $Q_{1}Q_{2}Q_{3}l_{2}$ below
the experimental limit even for values of $<l_{2N}>$ much less than the
Planck scale. Of course one must look at all dimension 5 operators. In the
example just considered one may see from Table \ref{Table:trilinear} that
there are potentially large contributions involving the fields $%
l_{1,2,3,4,5,6}.$ Thus we obtain suppression of dimension five operators
only if the light leptons are dominantly in the $l_{7,8,9}$ and
$\overline{l}%
_{1,3}$ and $\overline{l}_{2}|_{H}$ directions. Whether this happens depends
sensitively on the relative magnitude of the large vevs and on the
intermediate scale breaking and we will not consider this further here. 

\subsubsection{R-parity suppression}

Our discussion of the magnitude of the coefficient of the operator $%
Q_{1}Q_{2}Q_{3}l_{2}$ assumed that we could ignore the effect of Kaluza
Klein and string excitiations. This is not true when we have large vevs
developing along flat directions because the states light in the absence of
such vevs may acquire mass by mixing with the massive excitations. For
example, there may be a coupling $Q_{2}X_{q}l_{8}$ where $X_{q}$ is a Kaluza
Klein or string state with the gauge quantum numbers of $q_{i}$. Since
$X_{q}
$ must have a term in the effective Lagrangian $M_{X}X_{q}\overline{X}_{Q}$
the net effect of the large vev, $<l_{8}|_{\nu _{R}}>$, is to generate a mass
for the component proportional to $(<l_{8}|_{\nu
_{R}}>Q_{2}|_{3}+M_{X}X_{Q}|_{2})$. If $<l_{8}|_{\nu _{R}}>$ is larger than
$M_{X}$ this will be principally along the $Q_{2}$ direction showing this
mixing term is important when determining the mass of the ``light'' states.
One may now see that in the limit of large $<l_{8}|_{\nu _{R}}>$ there is a
suppression of dimension five contributions to nucleon decay due to the
effective $R-$parity discussed above. This follows because the quark and
lepton states are $C-$even while $l_{8}$ is $C-$odd. Thus one sees that the
term $\frac{1}{M^{2}}{\bf 27.27.27.27.<27}^{\dagger }{\bf >|}_{F}$
responsible for nucleon decay involves only $C-$even states in the
numerator. For the particular example discussed above the ${\bf <27}%
^{\dagger }{\bf >}$ vev is the $C-$even vev $<l_{2N}^{\dagger }>.$ However,
as just discussed, the mass, $M$, associated with this term is that of the
third component of $Q_{2}$ and is given by $M\approx $ $<l_{8}|_{\nu
_{R}}>.$
As a result the net contribution has the effective mass scale given by $%
M_{eff}^{-1}=\frac{{\bf <27}^{\dagger }{\bf >}}{M^{2}}\approx (\frac{%
<l_{2N}^{\dagger }>}{<l_{8}|_{\nu _{R}}>})^{2}\frac{1}{<l_{2N}^{\dagger
}>}.$
We see that relative to our previous estimate there is a suppression factor
$%
(\frac{<l_{2N}^{\dagger }>}{<l_{8}|_{\nu _{R}}>})^{2}$ which, for large $%
<l_{8}|_{\nu _{R}}>,$ can explain the needed suppression. Of course one must
consider too the contribution associated with the ``light'' component
proportional to $(M_{X}Q_{2}|_{3}-<l_{8}|_{\nu _{R}}>X_{Q}|_{2})$ which is
orthogonal to the heavy state proportional to $(<l_{8}|_{\nu
_{R}}>Q_{2}|_{3}+M_{X}X_{Q}|_{2}).$ It is straightforward to see that this
acquires mass of $O(<l_{2N}>)$ and could apparently give a larger $%
M_{eff}^{-1}.$ However, for $M_{X}\approx <l_{2N}^{\dagger }>$ it gives a
contribution of the same order as the one just discussed because the
coupling to the quark and lepton states is suppressed by the square of a
mixing angle of $O(\frac{M_{X}}{<l_{8}|_{\nu _{R}}>})$ in the large $%
<l_{8}|_{\nu _{R}}>$ limit. Thus we see that the residual approximate $R-$%
parity of the theory offers the possibility of an elegant explanation of the
suppression of dimension five contributions to nucleon decay. We will return
to a more complete discussion of this possibility elsewhere.

\section{Quark and Lepton masses}

We conclude with a discussion of the form of the quark and lepton masses
that result in the theory just discussed with a residual baryon parity. We
start with a discussion of the lepton mass matrix. The structure of the mass
matrix is constrained by the residual $Z_{3}^{G}$ symmetry. This means the
left-handed charge conjugate leptons ($SU(2)$ singlets) are mixtures of the
positive chirality states \ with $Z_{3}^{G}$ charges $0,4.$ In fact there is
a separate conservation of chirality for the two charges so from Table \ref
{Table:1} we see there are left light two states with charge $0$ and one
state with charge $4.$ The left-handed lepton doublets have a similar
structure but in this case the terms $0.2.2$, $0.0.4$ and $4.4.2$ induce
mixing of the ``lepton'' states $4_{l}$ and the ``Higgs'' states $\overline{4%
}_{H}$ and a mixing of the lepton state $0_{l}$ with the lepton state in $%
2_{l}$ and $\overline{2}_{l}.$ The light Higgs are the positive chirality
states of Table \ref{Table:1}, a mixture of the states $0_{H}$, $\overline{4}%
_{H}$ and $4_{l}.$ The resulting mass matrix has the form 
\begin{equation}
\left( 
\begin{array}{ccc}
0_{E} & 0_{E} & 4_{E}
\end{array}
\right) .\left( 
\begin{array}{ccc}
0 & 0 & <0_{H}> \\ 
0 & 0 & <0_{H}> \\ 
<0_{H}> & <0_{H}> & 0
\end{array}
\right) .\left( 
\begin{array}{c}
0_{l}+2_{l}+\overline{2}_{l} \\ 
0_{l}+2_{l}+\overline{2}_{l} \\ 
4_{l}+\overline{4}_{H}
\end{array}
\right)
\end{equation}
where we have suppressed the Yukawa couplings and the mixing factors
determining the light states composition. Note that the L-R symmetry is
broken by the mixing in the $l$ doublet sector. This mass matrix has two
non-zero eigenvalues. Their values depend on the precise nature of the
intermediate scale breaking and on the Yukawa couplings in both the
low-energy sector and the sector involving compactification scale masses. We
will explore this for a specific intermediate breaking scheme elsewhere.
Here we consider the more pressing problem whether the low-energy symmetries
allow for an electron mass. The masslessness of this state at tree level is
ensured by the $Z_{3}^{G}$ symmetry. At radiative order mass is generated by
the term $0.\overline{4}_{H}^{\dagger }$.$0|_{F}$ but SUSY\ protection means
that this is generated by SUSY breaking through the $D-$term $0.\overline{4}%
_{H}^{\dagger }$.$0.X^{\dagger }|_{D}=0.\overline{4}_{H}^{\dagger }$.$%
0|_{F}X^{\dagger }|_{\overline{F}}$ leading to the suppression factor of $O(%
\frac{h^{2}}{4\pi }.\frac{m_{3/2}}{M})$ where $h$ is the appropriate
combination of the various Yukawa couplings involved. This is too small to
be identified with the ratio $\frac{m_{e}}{m_{\tau }}$ and so this radiative
term is not able to generate an acceptable electron mass. However the $%
Z_{3}^{G}$ symmetry may be broken to give an electron mass without losing
the $Z_{3}^{G}$ protection for the light Higgs states or nucleon stability.
This is because we may induce mixing in the light lepton states via the
couplings $0_{E,l}.\overline{0}_{E,l}.<\phi _{0}>$ $\ $and $4_{E,l}.%
\overline{0}_{E,l}.<\phi _{-4}>$ where we have allowed a Standard Model
singlet field, $\phi ,$ with $Z_{2}^{G}$ charge $-4$ to acquire a vev. In
this case the mass matrix becomes

\begin{eqnarray}
&&\left( 
\begin{array}{ccc}
0_{E}+4_{E} & 0_{E}+4_{E} & 4_{E}+0_{E}
\end{array}
\right) .\left( 
\begin{array}{ccc}
<0_{H}>\eta & <0_{H}>\eta & <0_{H}> \\ 
<0_{H}>\eta & <0_{H}>\eta & <0_{H}> \\ 
<0_{H}> & <0_{H}> & <0_{H}>\eta
\end{array}
\right) .  \nonumber \\
&&\hspace{3.6in}.\left( 
\begin{array}{c}
0_{l}+2_{l}+\overline{2}_{l}+4_{l}+\overline{4}_{H} \\ 
0_{l}+2_{l}+\overline{2}_{l}+4_{l}+\overline{4}_{H} \\ 
4_{l}+\overline{4}_{H}+0_{l}+2_{l}+\overline{2}_{l}
\end{array}
\right)  \label{leptonmasses}
\end{eqnarray}
where 
\begin{equation}
\eta =\frac{<\phi _{-4}><\phi _{0}>}{<\phi _{-4}>^{2}+<\phi _{0}>^{2}}
\end{equation}
Clearly the electron now acquires a mass and its smallness may be explained
if $\frac{<\phi _{-4}>}{<\phi _{0}>}$ is small. This provides an
illustration how the string symmetries can generate an hierarchical
structure in the lepton masses. In this case the {\it same} symmetry
responsible for light Higgs states and for nucleon stability is responsible
for the electron mass hierarchy! One sees that a detailed prediction for the
lepton masses requires determination of a complicated mixing pattern amongst
the leptons and the Higgs, but this is characteristic of all attempts to
explain the fermion mass structure through Froggatt Nielsen mixing \cite
{froggatt}.

Of course it is necessary to check that the vev $<\phi _{-4}>$ does not lead
to large masses for the light Higgs doublets. To see that this is indeed the
case note that the constraints of holomorphicity of the superpotential
require that the only mass terms in the Higgs sector generated by $<\phi
_{-4}>$ come from the terms $\overline{0}.4_{H}.<\phi _{-4}>$ and $\overline{%
0}.\overline{0}.<\overline{0}>.<\phi _{-4}>.$ This only involves odd $%
Z_{2}^{G}$ chirality fields, all of which acquire large compactification
scale masses by coupling to even $Z_{2}^{G}$ chirality fields. Adding a
``Majorana'' type mass to the odd chirality states only does not affect the
light even chirality sector because there is no mass mixing of these states
with the the odd chirality fields. Of course at radiative order the light
even chirality sector {\it will} mix with the heavy sector but, just as in
the electron mass case discussed above, the resultant mass generated is
negligibly small. Similarly it is straightforward to verify that the new
terms do not spoil the approximate $R-$parity needed to stabilise the
nucleon. The reason is the same; the new couplings involve the fields
descending from the $\overline{27}$ and these are all odd parity quark
fields. The light quarks have even parity and do not couple to these new
couplings.

We postpone to another publication a detailed discussion of the form of the
neutrino mass structure. The Dirac mass matrix has a similar form to the
charged lepton mass matrix given above.The Majorana mass matrix for the
right handed fields comes from couplings of the form 
\begin{equation}
4_{{\overline{\nu }}_{R}}.0_{{\overline{\nu }}_{R}}.(-2)_{{\overline{H}}%
^{0}}.(-2)_{{\overline{H}}^{0}}
\end{equation}
In addition there are mass terms coming from the mixing of the right handed
neutrinos with singlet fields from the couplings of the form given in Table 
\ref{Table:trilinear1}. For example we have

\begin{equation}
<{\ l_{8}|_{\nu _{R}}>{\overline{l}}_{1}|_{\overline{\nu }_{R}}\phi _{18}}
\end{equation}
together with another seven such couplings. This gives a complicated mass
matrix the analysis of which lies beyond the scope of this paper.

We turn now to the quark masses. Consider the case that the Wilson line
breaking is associated with the cyclic permutation group. The vectorlike
pairs of up quarks acquire mass due to the large-scale breaking only via $%
{\bf 1.27.\overline{27}}$ related couplings. As a result the light quark states
involve mixing only between $0_{q,Q}$ and $4_{q,Q}$ respectively giving the
mass matrix of the form

\begin{eqnarray}
&&\left( 
\begin{array}{ccc}
0_{U}+4_{U} & 0_{U}+4_{U} & 4_{U}+0_{U}
\end{array}
\right) .  \nonumber \\
&&.\left( 
\begin{array}{ccc}
<0_{H}>\eta & <0_{H}>\eta & <0_{H}> \\ 
<0_{H}>\eta & <0_{H}>\eta & <0_{H}> \\ 
<0_{H}> & <0_{H}> & <0_{H}>\eta
\end{array}
\right) \allowbreak .\left( 
\begin{array}{c}
0_{q}+4_{q} \\ 
0_{q}+4_{q} \\ 
4_{q}+0_{q}
\end{array}
\right)  \label{up}
\end{eqnarray}
\newline
Note that this structure {\it preserves} the underlying $L-R$ symmetry
because there is no mixing with $\overline{2}$ in the quark sector$.$ For
this reason the case with $\frac{<\phi _{-4}>}{<\phi _{0}>}<<1$ is untenable
because it leads to two massive degenerate quark states and one light state.
As we note below this case may be rescued if we consider the case the Wilson
line breaking is associated with the phase twist generated by the (0,3,6,0)
element. First however we note that it is possible to live with this form of
the up quark matrix in the limit $\frac{<\phi _{-4}>}{<\phi _{0}>}>>1$. As
we \ noted above a large $<\phi _{-4}>$ does not spoil the light Higgs
protection or destabilise the nucleon so this limit is acceptable. Depending
on the details of the intermediate scale breaking and the Yukawa couplings
the diagonal entries in the mass matrix may be large so the symmetric
structure of the matrix may be acceptable. We will discuss this possibility
in detail elsewhere. In eq(\ref{leptonmasses}) one may see the lepton mass
matrix is also acceptable in this limit although the smallness of the
electron mass no longer follows simply from the magnitude of $\frac{<\phi
_{-4}>}{<\phi _{0}>}.$

Finally we consider the down quark masses. Its structure is similar to the
leptons due to the mixing in the down quark sector generated by the
${\bf 27}^{3}$
and ${\bf \overline{27}}^{3}$ terms. We find the mass matrix has the form

\begin{eqnarray}
&&\left( 
\begin{array}{ccc}
0_{D}+4_{D}+\overline{2}_{D} & 0_{D}+4_{D}+\overline{2}_{D} & 4_{D}+0_{D}+%
\overline{2}_{D}
\end{array}
\right) .  \nonumber \\
&&.\left( 
\begin{array}{ccc}
<0_{H}>\eta & <0_{H}>\eta & <0_{H}> \\ 
<0_{H}>\eta & <0_{H}>\eta & <0_{H}> \\ 
<0_{H}> & <0_{H}> & <0_{H}>\eta
\end{array}
\right) .\left( 
\begin{array}{c}
0_{d}+4_{l} \\ 
0_{l}+4_{l} \\ 
4_{l}+0_{l}
\end{array}
\right)  \label{down}
\end{eqnarray}
\linebreak Again this has an acceptable structure although the hierarchy of
masses is not explained in the limit $\eta >>1.$ Note that the down quark
and lepton masses may still be related by the underlying $E_{6}$ symmetry
despite the fact that the gauge group after compactification is $%
SU(3)\otimes SU(2)\otimes U(1).$ The reason is that the quark and lepton
fields are made up of the same $E_{6}$ multiplets but with different phases
associated with the components. Thus the lepton multiplets, being
permutation singlets, have the form $A+B+C$ where $A\rightarrow $ $B$ $%
\rightarrow $ $C$ $\rightarrow A$ under the permutation symmetry. The quark
doublets have the form $A+\alpha ^{2}B$ +$\alpha C$ where $\alpha =e^{2\pi
i/3}$. As a result the couplings in the compactified model are related to
the same underlying $E_{6}$ couplings, albeit modified by an overall phase.
We will discuss in another paper whether this phase can be responsible for
the observed CP\ violation in the Standard Model.

To complete this section we consider the structure of quark masses for the
case the Wilson line breaking is associated with the phase twist generated
by the (0,3,6,0) element. This illustrates some of the diversity associated
with different string constructions and also gives an example in which the
case $\eta <<1$ is tenable and some of the fermion mass hierarchy is
determined by the {\it same} symmetry responsible for light Higgs states and
for nucleon stability. In this case there are only three generations of
quarks on compactification and no vectorlike representations. As a result
the structure of the quark mass matrix is simplified as there is no mixing
to consider when determining the light spectrum. From Table \ref{Table:4}
and Table \ref{Table:trilinear} we find the following form for the quark
mass matrices 
\begin{equation}
\left( 
\begin{array}{ccc}
Q_{1} & Q_{2} & Q_{3}
\end{array}
\right) .\left( 
\begin{array}{ccc}
0 & l_{5} & l_{4,6} \\ 
l_{4} & l_{2} & l_{1} \\ 
l_{5,6} & l_{1} & l_{2,3}
\end{array}
\right) .\left( 
\begin{array}{c}
q_{1} \\ 
q_{2} \\ 
q_{3}
\end{array}
\right)  \label{qmm}
\end{equation}
From the discussion above we know that the light Higgs are mixtures of the $%
Z_{3}^{G}$ charge 0,$\overline{2},\overline{4}$ states. Only the $0$ states
are relevant here, namely $l_{1,2,4,5,7}.$ From eq(\ref{qmm}) we see that a
Higgs with these components this will generate an acceptable mass matrix
although the hierarchy of the second and third generations will require a
cancellation between the terms involving $l_{1,2}.$ We shall discuss ways in
which this may happen naturally in a separate paper. As for the leptons the
relative magnitude of the first generation masses is set by the $Z_{3}^{G}$
symmetry breaking structure via the ratio $\frac{<l_{4,5}>}{<l_{1,2}>}.$ A
particularly interesting feature is the appearance of the texture zero in
the (1,1) position. Such texture zeros are known to be consistent with the
observed pattern of fermion masses and if, in addition, the (1,2) entry
equals the (2,1) entry one obtains a {\it prediction} for the Cabibbo angle
which is known to be in excellent agreement with experiment \cite{weinberg}.
It is of interest to note that the symmetries of the string model lead
readily to such a symmetric structure. This follows if $Z_{2}$ associated
with anticyclic permutations of the three level $16$ factors is unbroken. In
this case the Higgs can only involve the $Z_{2}\,$\ even combination $%
l_{4+}=l_{4}+l_{5}$ leading to the mass matrix

\begin{equation}
\left( 
\begin{array}{ccc}
Q_{1} & Q_{2} & Q_{3}
\end{array}
\right) .\left( 
\begin{array}{ccc}
0 & l_{4+} & l_{4+} \\ 
l_{4+} & l_{2} & l_{1} \\ 
l_{4+} & l_{1} & l_{2,3}
\end{array}
\right) .\left( 
\begin{array}{c}
q_{1} \\ 
q_{2} \\ 
q_{3}
\end{array}
\right)
\end{equation}
This structure reproduces the good texture zero prediction for the CKM\
matrix given by \cite{weinberg}

\begin{eqnarray}
|V_{us}| &=&|\sqrt{\frac{M_{D}}{M_{s}}}\,\,-\sqrt{\frac{M_{u}}{M_{c}}}%
\,\,e^{i\sigma }|  \label{eq17} \\
c.f.(0.218-0.224) &=&|(0.16-0.33)-(0.047-0.07)e^{i\sigma }|  \nonumber
\end{eqnarray}
As a bonus the matrix has relatively small entries in the (3,1) and (1,3)
positions because, with the Higgs having the appropriate $l_{4+}$ magnitude
to give the correct (1,2) element, one finds that the (3,1) and (1,3)
elements do not affect the masses  to leading order. They can affect
the mixing angles but one may check that they give the relations

\begin{eqnarray}
|V_{ub}| & \simeq &\sqrt{\frac{M_{u}}{M_{c}}}\,\,|V_{cb}|  \label{eq14} \\
|V_{tu}| & \simeq
&\sqrt{\frac{M_{d}}{M_{s}}}\sqrt{\frac{M_{c}}{M_{u}}}\,\,|V_{ub}|,
\nonumber
\end{eqnarray}
where the $\simeq$ means equality up to coefficients of $O(1)$. These
relations
are also in excellent agreement with experiment. Thus we see that
texture zeros and a hierarchical structure can follow very naturally
from the underlying  string
symmetries of the three generation Calabi-Yau model.

\section{Summary and Conclusions}

In this paper we have explored a new class of string compactifications
obtained by allowing vevs to develop along absolutely flat directions. In
practice these directions may only be recognised starting with a four
dimensional string theory at a point in moduli space with enhanced symmetry.
In this discrete symmetries play a crucial role as only they are capable of
ensuring flatness to all non-renormalisable orders in the superpotential. It
turns out that the Gepner construction which builds a four dimensional
theory via products of $N=2$ superconformal theories has such R symmetries
together with additional gauge and discrete symmetries. It is known that
these compactifications are equivalent in the large radius limit to Calabi
Yau compactification and thus are of relevance to M-theory compactification.

As an example of the method we studied the flat directions in a three
generation Calabi Yau model built using the Gepner construction. The
original theory has an underlying $E_{6}$ gauge group which may be broken by
Wilson lines to $SU(3)^{3}$ but not to the Standard Model. The hope is that
the perturbation along flat directions will allow for a reduction of the
original symmetry to $SU(5)$ so that after Wilson line breaking the group
can be reduced to $SU(3)\otimes SU(2)\otimes U(1).$ Amongst other things
this relates the compactification scale determined in M-theory with the
gauge unification scale found by continuing the couplings of the MSSM and
offers the prospect of a quantitative string theory prediction for this
scale. We found that the theory does indeed have flat directions which lead
to a reduction of the string gauge symmetry to $SU(5).$

Construction of a phenomenological string theory requires that there be low
energy symmetries which stabilise the nucleon and also leave just one pair
of Higgs doublets light. We showed that the three generation Calabi Yau
model has just such a symmetry following from the original discrete
symmetries of the string. Furthermore we showed that the flat direction
leading to the reduced gauge symmetry leaves this symmetry unbroken so that
is possible to build a viable low energy theory which has just the Standard
Model $SU(3)\otimes SU(2)\otimes U(1)$ gauge symmetry below the
compactification scale. The stability of the nucleon follows from a ``Baryon
parity'' which forbids rapid baryon number violating processes but allows
lepton number processes to occur through the exchange of particles with
supersymmetry breaking scale masses. The stability of just one pair of Higgs
doublets follows through a new ``Higgs chirality'' which has an excess of
one positive chirality Higgs field of each weak hypercharge. Remarkably the
stability of the nucleon and the lightness of just one pair of Higgs
doublets is ensured by the same $Z_{3}^{G}$ discrete symmetry. A feature of
the theory at the compactification scale is that there are several
additional states left light in vectorlike representations with respect to $%
SU(3)\otimes SU(2)\otimes U(1).$ We show that these states are all likely to
acquire intermediate scale masses somewhere between the compactification and
the electroweak scale. This is generated by intermediate scale vevs
triggered by $SU(3)\otimes SU(2)\otimes U(1)\ $\ scalar fields acquiring
negative masses squared due to radiative corrections. We studied the
possibility for such intermediate scale breaking and showed that it is
likely to occur at a very high scale due to the constrained form of the
underlying string symmetry. We demonstrated that this can happen along
directions in field space which leave the $Z_{3}^{G}$ symmetry intact$.$

Given knowledge of the string symmetries it is straightforward to determine
the pattern of quark and lepton masses. In the three generation Calabi Yau
theory the residual low-energy symmetries impose constraints on the allowed
form of these matrices. We showed that an acceptable pattern of lepton
masses is possible in which the lightness of the electron may be determined
by the ratio of $Z_{3}^{G}$ symmetry breaking vevs. We studied the quark
masses in two variants of the theory in which the Wilson line breaking is
associated with different discrete groups which are modded out in the
construction of the three generation theory. In both cases an acceptable
pattern of masses for the up and down quarks is consistent with the
low-energy symmetries.

In the case the Wilson line breaking is associated with associated with the
cyclic permutation group one loses the association of the electron mass
hierarchy with the ratio of $Z_{3}^{G}$ symmetry breaking vevs. However one
finds in this case that the quark and lepton Yukawa couplings responsible
for generating masses have a residual Grand Unified structure associated
with the underlying $E_{6}$ which may preserve the successful $SO(10)$
relations between the charged lepton and down quark masses.

In the case the Wilson line breaking is associated with the phase twist
generated by the (0,3,6,0) element the quark mass structure is consistent
with the electron mass hierarchy being related to the $Z_{3}^{G}$ symmetry
breaking and in addition the lightness of the first generation of quarks is
also determined by the ratio of $Z_{3}^{G}$ symmetry breaking vevs. Moreover
we found that a symmetric mass matrix and texture zeros in the (1,1) and
(1,3) positions follow quite readily from the symmetries of the theory. This
structure leads to excellent predictions for the quark masses.

To go further it is necessary to study a specific pattern of intermediate
scale breaking. A given pattern will have further structure in the quark and
lepton masses following from the underlying rich structure of string
symmetries. We will report elsewhere on a study of a particularly promising
choice for this intermediate scale breaking.

W.P. would like to thank M.G. Schmidt and C.M.A. Scheich for
discussions.

\end{document}